\documentclass[10pt,letterpaper]{article}
\usepackage[top=0.85in,left=2.75in,footskip=0.75in]{geometry}

\usepackage{amsmath,amssymb}

\usepackage{changepage}

\usepackage[utf8x]{inputenc}

\usepackage{textcomp,marvosym}

\usepackage{cite}

\usepackage{nameref,hyperref}

\usepackage[right]{lineno}

\usepackage{microtype}
\DisableLigatures[f]{encoding = *, family = * }

\usepackage[table,dvipsnames]{xcolor}

\usepackage{array}

\newcolumntype{+}{!{\vrule width 2pt}}

\newlength\savedwidth

\raggedright
\usepackage[aboveskip=1pt,labelfont=bf,labelsep=period,justification=raggedright,singlelinecheck=off]{caption}

\makeatletter
\renewcommand{\@biblabel}[1]{\quad#1.}
\makeatother

\usepackage{lastpage,fancyhdr,graphicx}
\usepackage{epstopdf}
\fancyheadoffset[L]{2.25in}
\fancyfootoffset[L]{2.25in}
\newcommand{\taum}{\ensuremath{\tau_\mathrm{m}}}
\newcommand{\taud}{\ensuremath{\tau_\mathrm{d}}}
\newcommand{\tauf}{\ensuremath{\tau_\mathrm{f}}}

\newcommand{\Jei}{\ensuremath{J_\mathrm{ei}}}
\newcommand{\Jie}{\ensuremath{J_\mathrm{ie}}}
\newcommand{\Jii}{\ensuremath{J_\mathrm{ii}}}
\newcommand{\Jeep}{\ensuremath{J_\mathrm{ee}^\mathrm{(s)}}}
\newcommand{\Jeeb}{\ensuremath{J_\mathrm{ee}^\mathrm{(c)}}}

\newcommand{\Npop}{\ensuremath{N_\mathrm{pop}}}

\newcommand{\Iback}{\ensuremath{I_\mathrm{B}}}
\newcommand{\Istim}{\ensuremath{I_\mathrm{S}}}

\newcommand{\dd}{\ensuremath{\mathrm{d}}}

\newcommand{\Vp}{\ensuremath{V_\mathrm{p}}}
\newcommand{\Vr}{\ensuremath{V_\mathrm{r}}}

\newcommand{\red}[1]{\textcolor{black}{#1}}
\newcommand{\blue}[1]{\textcolor{blue}{#1}}

\newcommand{\be}{\begin{equation}}
\newcommand{\ee}{\end{equation}}

\usepackage{hyperref}
\definecolor{hypcolor}{named}{Blue}
\hypersetup{colorlinks=true, citecolor=hypcolor, urlcolor=hypcolor, linkcolor=hypcolor}

\usepackage{upgreek}
\begin{document}
\vspace*{0.2in}

\begin{flushleft}
{\Large
\textbf\newline{Exact neural mass model for synaptic-based working memory} % Please use "sentence case" for title and headings (capitalize only the first word in a title (or heading), the first word in a subtitle (or subheading), and any proper nouns).
}
\newline
%% Insert author names, affiliations and corresponding author email (do not include titles, positions, or degrees).
%\\
Halgurd Taher\textsuperscript{1},
Alessandro Torcini\textsuperscript{2,3},
Simona Olmi\textsuperscript{1,3,*},
%with the Lorem Ipsum Consortium\textsuperscript{\textpilcrow}
\\
\bigskip
\textbf{1} Inria Sophia Antipolis M\'{e}diterran\'{e}e Research Centre, MathNeuro Team, 2004 route des Lucioles, Bo\^{\i}te Postale 93. 06902 Sophia Antipolis cedex, France
\\
\textbf{2} Laboratoire de Physique Th\'eorique et Mod\'elisation, Universit\'e de Cergy-Pontoise,CNRS, UMR 8089, 95302 Cergy-Pontoise cedex, France 
\\
\textbf{3} CNR - Consiglio Nazionale delle Ricerche - Istituto dei Sistemi Complessi, 50019, Sesto Fiorentino, Italy
\\
\bigskip
%
%% Insert additional author notes using the symbols described below. Insert symbol callouts after author names as necessary.
%% 
%% Remove or comment out the author notes below if they aren't used.
%%
%% Primary Equal Contribution Note
%\Yinyang These authors contributed equally to this work.
%
%% Additional Equal Contribution Note
%% Also use this double-dagger symbol for special authorship notes, such as senior authorship.
%\ddag These authors also contributed equally to this work.
%
%% Current address notes
%\textcurrency Current Address: Dept/Program/Center, Institution Name, City, State, Country % change symbol to "\textcurrency a" if more than one current address note
%% \textcurrency b Insert second current address 
%% \textcurrency c Insert third current address
%
%% Deceased author note
%% Group/Consortium Author Note
%\textpilcrow Membership list can be found in the Acknowledgments section.
%
%% Use the asterisk to denote corresponding authorship and provide email address in note below.
* simona.olmi@inria.fr

\end{flushleft}

\section*{Abstract}
\red{A synaptic theory of Working Memory (WM) has been developed in the last
decade as a possible alternative to the persistent spiking paradigm.
In this context, we have developed a neural mass model able to reproduce
exactly the dynamics of heterogeneous spiking neural networks
encompassing realistic cellular mechanisms for short-term synaptic plasticity.
This population model reproduces the macroscopic dynamics
of the network in terms of the firing rate and the mean membrane potential.
The latter quantity allows us to get insigth on Local Field Potential and electroencephalographic
signals measured during WM tasks to characterize the brain activity.
More specifically synaptic facilitation and depression integrate
each other to efficiently mimic WM operations via either synaptic reactivation
or persistent activity.  Memory access and loading are associated to stimulus-locked
transient oscillations followed by a steady-state activity in the
$\beta$-$\gamma$ band, thus resembling what observed in the cortex
during vibrotactile stimuli in humans and object recognition in monkeys.
Memory juggling and competition emerge already by loading only two items. However
more items can be stored in WM by considering neural architectures
composed of multiple excitatory populations and a common inhibitory pool.
Memory capacity depends strongly on the presentation rate of the items and it maximizes for an optimal frequency range.
In particular we provide an analytic expression for the maximal memory capacity.
Furthermore, the mean membrane potential turns out to be
a suitable proxy to measure the memory load, analogously to
event driven potentials in experiments on humans.
Finally we show that the $\gamma$ power increases with the number of loaded items,
as reported in many experiments, while $\theta$ and $\beta$
power reveal non monotonic behaviours.
In particular, $\beta$ and $\gamma$ rhytms are crucially sustained by the
inhibitory activity, while the $\theta$ rhythm is controlled by excitatory
synapses. 
%Finally, an analytic expression for the maximal capacity is obtained.
}
% Please keep the Author Summary between 150 and 200 words
% Use first person. PLOS ONE authors please skip this step. 
% Author Summary not valid for PLOS ONE submissions.   

\section*{Author summary}
\red{Working Memory (WM) is the ability to temporarily store and manipulate stimuli
representations that are no longer available to the senses.
We have developed an innovative coarse-grained population model able to mimic several operations
associated to WM. The novelty of the model consists in reproducing exactly the dynamics
of spiking neural networks with realistic synaptic plasticity composed of hundreds
of thousands neurons in terms of a few macroscopic variables.
These variables give access to experimentally measurable quantities such
as local field potentials and electroencephalografic signals.
Memory operations are joined to sustained or transient oscillations emerging
in different frequency bands, in accordance with experimental results for primate
and humans performing WM tasks. We have designed an architecture composed
of many excitatory populations and a common inhibitory pool able to
store and retain several memory items. The capacity of our multi-item architecture
is around 3-5 items, a value corresponding to the WM capacities measured in many experiments.
Furthermore, the maximal capacity is achievable only for presentation rates within
an optimal frequency range. Finally, we have defined a measure of the memory load
analogous to the event-related potentials employed to test humans' WM capacity during visual memory tasks.}
%\linenumbers

% Use "Eq" instead of "Equation" for equation citations.
\section*{Introduction}

Working memory (WM) allows us to keep recently accessed information, available for manipulation:
it is fundamental in order to pass from reflexive input-output reactions to the organization of goal-directed behaviour \cite{just1992capacity, goldman1995cellular, engle1999working, fuster1999memory, vogel2004neural, chatham2015multiple}. Starting from the seminal work of Fuster and Alexander \cite{fuster1971neuron}, several experiments have shown that neurons in higher-order cortex, including the prefrontal cortex (PFC), are characterized by elevated levels of spiking activity during memory delays related to WM tasks.  Experimental results seem to suggest that WM maintenance is enhanced by delayed spiking activity, engaging executive functions associated with large part of the cortex, from frontal to posterior cortical areas \cite{miller2001integrative,lara2015role}.
\red{In particular, it has been shown that neurons in the PFC exhibit persistent activity selective to the sample cue during oculomotor delayed-response task \cite{funahashi1989mnemonic} as well as during vibrotactile response task \cite{romo1999} and that this activity is robust to distractors \cite{miller1996neural}.}
\red{Therefore, classic models proposed persistent spiking for online maintenance of information, since neural ensembles in an active state appear more prone to fast processing 
\cite{dipoppa2016, constantinidis2018}. These models have been able to describe multi-item loading and maintenance in WM \cite{amit1997}, spatial WM in the cortex \cite{compte2000synaptic}, and two interval discrimination tasks \cite{machens2005}.}

\red{A further relevant aspect, associated to WM operations, is the presence of neural oscillations, whose increase in the oscillatory power during WM maintenance and rehearsal has been reported for humans
in the $\theta$-band (4-8 Hz) \cite{gevins1997,jensen2002}, as well as in the $\beta$ (12-25 Hz) and $\gamma$-range (25-100 Hz) \cite{tallon1998,howard2003gamma}, while for monkeys it has been reported in the $\theta$ and $\gamma$-range \cite{pesaran2002,wimmer2016} joined to a decrease in the $\beta$-band \cite{wimmer2016}. 
The results for the activity in the $\alpha$-band (8-11 Hz) are somehow more complex: on one side it has been shown that during WM retention in humans no variations are observable \cite{tallon1998}
while, on the other side, that it can be associated to the inhibitory action suppressing irrelevant stimuli \cite{sauseng2009}.} \red{
Despite many studies, the role played by oscillations in WM is still unclear; however numerical studies have suggested that different cycles of a high frequency oscillation in the $\gamma$-band can encode a sequence of memory items if nested within a slower $\theta$ or $\beta$ rhythm \cite{lisman1995,jensen1996novel,kopell2011}. Moreover, it has been shown in computational models that oscillatory forcing in different frequency bands may provide effective mechanisms for controlling the persistent activity of neural populations and hence the execution of WM tasks \cite{di2013,schmidt2018network}.
}

\red{Recently, the persistent state paradigm has been criticized on the basis of the inconsistencies emerging in data processing: persistent activity is the result of specific data processings, being neural spiking averaged over time and across trials.} In particular, averaging across trials can create the appearance of persistent spiking even when, on individual trials, spiking is actually sparse \cite{shafi2007variability, lundqvist2016gamma}. Anyhow there are examples of single neurons that show real persistent spiking on individual trial rasters, but they represent a small percentage, while the bulk of neurons spike sparsely during WM delays \cite{hussar2012memory, fujisawa2008behavior}.

A pioneering study \cite{wang2006heterogeneity} revealed that the interactions among pyramidal neurons in the PFC display
synaptic facilitation lasting hundreds of milliseconds. This study paved the way for a development of an alternative model for WM based on synaptic features. In particular, memory items can be stored by spiking-induced changes in the synaptic weights and these items can be maintained in WM by brief bursts of spiking restoring the synaptic strengths \cite{mongillo2008synaptic, lundqvist2011theta, rose2016, miller2018working}. \red{As a result} this kind of model is more justifiable from a metabolic point of view, since the memory requires less resources to be held with respect to the persistent activity. \red{Moreover}, multiple memory items can be maintained in WM at the same time by having different neuronal populations emitting short bursts at different times. This would solve problems usually observables in models based on persistent spiking: namely, the interference among different stored items and the disruption of memories associated to new sensory inputs \cite{siegel2009phase, miller2018working}.

\red{
In this context, a fundamental model for WM based on short-term synaptic plasticity (STP) 
has been introduced by Mongillo {\it et al.} in \cite{mongillo2008synaptic}.
In particular, in this model, synapses among pyramidal neurons display depressed and facilitated transmissions
based on realistic cellular mechanisms \cite{tsodyks1997neural,tsodyks1998neural,markram1998}. Synaptic facilitation allows the model to maintain an item stored for a certain period in WM, without the need of an enhanced spiking activity. Furthermore, synaptic depression is responsible for the emergence of population bursts (PBs), which correspond to a sub-population of neurons firing almost synchronously within a short time window \cite{tsodyks2000, luccioli2014}. This WM mechanism is implemented in \cite{mongillo2008synaptic} within a recurrent network of spiking neurons, while a 
simplified firing rate model is employed to gain some insight into the population dynamics. 
The rate model, like most of the models investigated in literature 
\cite{wilson1972excitatory, mi2017synaptic}, is heuristic, i.e. the macroscopic 
description had not a precise correspondence with the microscopic neuronal evolutions.}

\red{ 
Recent results for the primate PFC have revealed that the mnemonic stimuli are encoded at a population level allowing for a stable and robust WM maintenance, despite the complex and heterogeneous temporal dynamics displayed
by the single neurons \cite{murray2017}. This analysis suggests that the development of  
population models able to reproduce the macroscopic dynamics of 
heterogeneous spiking networks can be extremely useful to shed further light on the mechanisms at the basis
of WM. An ideal candidate to solve this task is represented by a neural mass model of new generation
able to exactly reproduce the macroscopic dynamics of spiking neural networks \cite{luke2013,laing2014,montbrio2015macroscopic}.
This exact derivation is possible for networks of  Quadratic Integrate and Fire (QIF) neurons, 
representing the normal form of Hodgkin's class I excitable membranes \cite{ermentrout1986parabolic}, thanks to the analytic techniques developed for coupled phase oscillators \cite{ott2008}. This new generation of neural mass models has been recently used to describe the emergence of collective oscillations (COs) in fully coupled networks \cite{devalle2017firing,laing2017,coombes2019,dumont2019}
as well as in balanced sparse networks \cite{volo2018}. Furthermore, it has been successfully employed
to reveal the mechanisms at the basis of theta-nested gamma oscillations \cite{segneri2020theta,ceni2020cross} 
and of the coexistence of slow and fast gamma oscillations \cite{bi2020}. 
However, to our knowledge, such models have not been yet generalized to spiking networks with plastic synapses.}

\red{ 
Our aim is to develop a next generation neural mass model encompassing short-term synaptic facilitation and depression \cite{tsodyks1998neural}. 
This model will enable us to revise the synaptic theory of working memory \cite{mongillo2008synaptic}
with a specific focus on the emergence of neural oscillations and their relevance for WM operations.
In particular, the newly derived neural mass model will capture the macroscopic evolution of the network 
not only in terms of the firing rate, as the standard heuristic models do \cite{wilson1972excitatory}, but also of the mean membrane potential \cite{montbrio2015macroscopic}. This will allow for a more direct 
comparison with the results of electrophysiological experiments often employed to characterize WM processes.
Indeed, as shown in \cite{okun2010}, electroencephalograms (EEGs),  event-related potentials (ERPs) and Local Field 
Potentials (LFPs) have the same information content as the mean membrane potentials.
}

\red{
The paper is organised as follows. Firstly, we will report clear evidence that the employed neural mass model reproduces,
with extreme accuracy, the macroscopic dynamics of heterogeneous networks of QIF neurons with STP.
Secondly, we will show that the model is able to mimic the fundamental operations required for WM functioning,
whenever memory items are maintained either via spontaneous and selective reactivation or via
persistent activity. In particular, we will devote particular attention to the spectral features of COs 
emerging during such operations and to their analogy with experimental findings reported for
EEG responses to vibrotactile stimuli in primary somatosensory cortex
in humans \cite{spitzer2010} and for LFP measurements performed in PFC of primates 
during objects coding in WM \cite{siegel2009phase,lundqvist2016gamma}. In this context, we will show that 
a heuristic firing rate model, despite being specifically designed to reproduce the mean-field dynamics of QIF networks, is unable to
sustain fast oscillations, which usually characterize the neural activity during WM tasks. 
Then we will perform a detailed analysis of the competition between two consecutively loaded memory items, 
whenever the memory maintenance is realised in terms of either synaptic facilitation or persistent spiking.
Moreover, we will analyse how the memory capacity depends on the presentation rate of
a sequence of items and which frequency bands are excited during loading and maintenance
of multiple items in WM. We will also derive an analytic estimation of the maximal working memory capacity
for our model, by extending recent results obtained in \cite{mi2017synaptic}. 
Furthermore, we will show that the mean membrane potential can represent a proxy to investigate
memory load and capacity, in analogy with a series of experiments on neurophysiological measures 
of the WM capacity in humans \cite{vogel2004neural,vogel2005neural}.
Finally, we will conclude with a discussion on the obtained results and their possible relevance for
the field of neuroscience.
}

\section*{Results}
\label{sec:results}

We develop a model for WM able to memorize discrete items by following \cite{amit1997model}, in particular we 
consider $\Npop$ coupled excitatory populations, each coding for one item, and a single inhibitory population connected
to all the excitatory neurons. This architecture is justified by recent experimental results indicating that 
GABAergic interneurons in mouse frontal cortex are not arranged in sub-populations and that they densely innervate all pyramidal cells \cite{fino2011}. The role of inhibition is to avoid abnormal synchronization 
and to allow for a competition of different items once stored in the excitatory population activity.
Furthermore, in order to mimic synaptic-based WM  we assume that only the excitatory-excitatory synapses are plastic displaying short-term depression and facilitation \cite{mongillo2008synaptic}.

The macroscopic activity of $\Npop$ excitatory and one inhibitory populations of heterogeneous QIF neurons can be exactly described in terms of the population firing rate $r_k(t)$ and mean membrane voltage $v_k(t)$ of each population $k$ by the following set of ODEs \cite{montbrio2015macroscopic}:
\begin{subequations}\label{eq:fre}
	\begin{align}
		\taum^{n} \dot{r}_k&=\frac{\Delta_k}{\taum^{n} \pi} + 2 r_k v_k \qquad k=0,1,\dots,\Npop \\
		\taum^{n} \dot{v}_k&=v_k^2 + H_k -(\pi \taum^{n} r_k)^2 +\Iback + \Istim^{(k)}(t)+ 
		\taum^{n} \sum_{l=0}^{\Npop} {\tilde J}_{kl}(t) r_l \quad ;
	\end{align}
\end{subequations} 
where $\taum^{n}$ denotes the membrane time constant of excitatory (inhibitory) populations $n=\mathrm{e}$ ($n=\mathrm{i}$)
and $\Iback$ is a constant background current common to all populations, whereas the (time dependent) stimulus current $\Istim^{(k)}(t)$ may be population specific. Excitatory populations are characterized by $k>0$, the inhibitory by $k=0$.
In absence of STP and for instantaneous synapses the synaptic weights are constant in time ${\tilde J}_{kl}(t) = J_{kl}$ and their sign determines whether the connections are excitatory ($J_{kl}>0$) or inhibitory ($J_{kl}<0$). 
The heterogeneous nature of the neurons is taken into account by considering randomly distributed neural excitabilities, reflecting their biological variability. In particular, we assumed that the distribution of the neural excitabilities is a Lorentzian characterized by a median $H_k$ and a half width half-maximum (HWHM)  $\Delta_k$ \footnote{\red{The single QIF 
neuron in absence of external and recurrent stimulations is supra-threshold (sub-threshold) if its excitability is positive (negative). Therefore a positive (negative) value of 
$H_k$ ensures that the majority of the neurons are supra-threshold (sub-threshold). However, due to the shape of 
the distribution, the population is always composed of both excitable and tonic firing neurons, the percentage of which
is determined by the values of $H_k$ and $\Delta_k$.}}.This choice allows us to derive the neural mass model from the spiking QIF network analytically. However, the overall dynamics does not change substantially by considering other distributions for the neural excitabilities, like Gaussian and Binomial ones \cite{montbrio2015macroscopic,bi2020}. 
 
Plasticity enables synapses to modify their efficacy in response to spiking activity and hence to stimuli. As opposed to long-term plasticity, STP occurs on short time scales in the order of hundred milliseconds to a few seconds. In particular one distinguishes between synaptic depression, i.e. the weakening, and facilitation, i.e. the strengthening of synapses. The phenomenological model for neocortical synapses, introduced by Tsodyks and Markram to reproduce the synaptic depression of pyramidal neurons \cite{tsodyks1997neural}, assumes that each synapse is characterized by a finite amount of resources:
each incoming pre-synaptic spike activates a fraction of resources, which rapidly inactivates in a few milliseconds and recover from synaptic depression over a longer time-scale $\taud$ \red{($\sim$ hundreds of milliseconds)}. The formulation of a facilitating mechanism has been subsequently developed by Tsodyks, Pawelzik and Markram \cite{tsodyks1998neural}: in this case, the fraction of resources is increased at each pre-synaptic spike and returns to the baseline value with a time constant \red{$\tauf\sim 1$ s}. The idea behind these phenomenological STP models is that the magnitude of post-synaptic potentials (PSPs) is weighted by the relative amount $X_i\in [0,1]$ of available pre-synaptic resources and by the fraction $U_i\in[0,1]$ of these resources utilized to generate the PSP. Whenever the neuron $i$ emits a spike, neurotransmitters are released and $X_i$ decreases, while at the same time calcium accumulates in the pre-synaptic terminals, increasing the neurotransmitter release probability at the next spike emission and hence the fraction $U_i$ of utilized resources. This form of STP is purely pre-synaptic, i.e., depression and facilitation are influenced only by spikes emitted by the pre-synaptic neuron, this justifies why we can assume that the state of all the efferent synapses of a neuron $i$ is described
by only the single couple of variables $(X_i,U_i)$ \cite{di2013}. 
Furthermore, it is fundamental that the facilitation timescale $\tauf$ is longer than that associated to the recover from depression $\taud$ \cite{mongillo2008synaptic}. In this way the information provided by the stimulus will be carried over a time $\tauf$ by the facilitated  synapse.

The inclusion of the depression and facilitation STP mechanisms  in the spiking networks
is reflected in the modification of the synaptic weights entering in the neural mass model (Eqs. \eqref{eq:fre}). 
In particular, these are rewritten as  ${\tilde J}_{kl}(t) = J_{kl} u_l(t) x_l(t)$ if $k$ and $l$ are
both excitatory populations, and simply as  ${\tilde J}_{kl}(t) = J_{kl} $ if either $k$ or $l$ is inhibitory.
The terms $x_k(t) $ and  $u_k(t)$ represent the mean available resources and the mean utilization factor of the population $k$, respectively. 
If we neglect the very fast inactivation of the depression terms, the dynamical evolution of  $x_k$ and  $u_k$ is regulated by the following ODEs
\begin{subequations}\label{eq:dumeanfied}
	\begin{align}
		\frac{\dd x_k(t)}{\dd t} &=\frac{1-  x_k(t)  }{\taud}-  u_k(t)   x_k(t)   r_k(t)\\	
		\frac{\dd u_k(t)}{\dd t} &=\frac{U_0-  u_k(t)  }{\tauf}+U_0(1-u_k(t))r_k(t) \qquad {k=1,\dots,\Npop}
		\quad ;
	\end{align}
\end{subequations} 
where $U_0=0.2$ is the baseline value of the utilization factor, while $\taud=200$ ms, $\tauf=1500$ ms represent the depression and facilitation time scales, respectively. These parameter values are fixed for all the simulations reported
in this paper. 

The model \mbox{\eqref{eq:fre} - \eqref{eq:dumeanfied}} describes the dynamics of one inhibitory 
and $\Npop$ excitatory coupled neuronal populations with STP in terms of their population firing rates $r_k$, mean membrane voltages $v_k$, mean available resources $x_k$ and mean utilization factors $u_k$. 
Details on the derivation of the neural mass model Eqs. \eqref{eq:fre} - \eqref{eq:dumeanfied} and on the underlying spiking networks can be found in the sub-sections \nameref{sec:nn}, \nameref{sec:nm}, \nameref{sec:mp} in \nameref{sec:methods}.

%\clearpage
\subsection*{Network dynamics versus neural mass evolution}
\label{sec:comparison}

\begin{figure}[!h]
\includegraphics[width=1\linewidth]{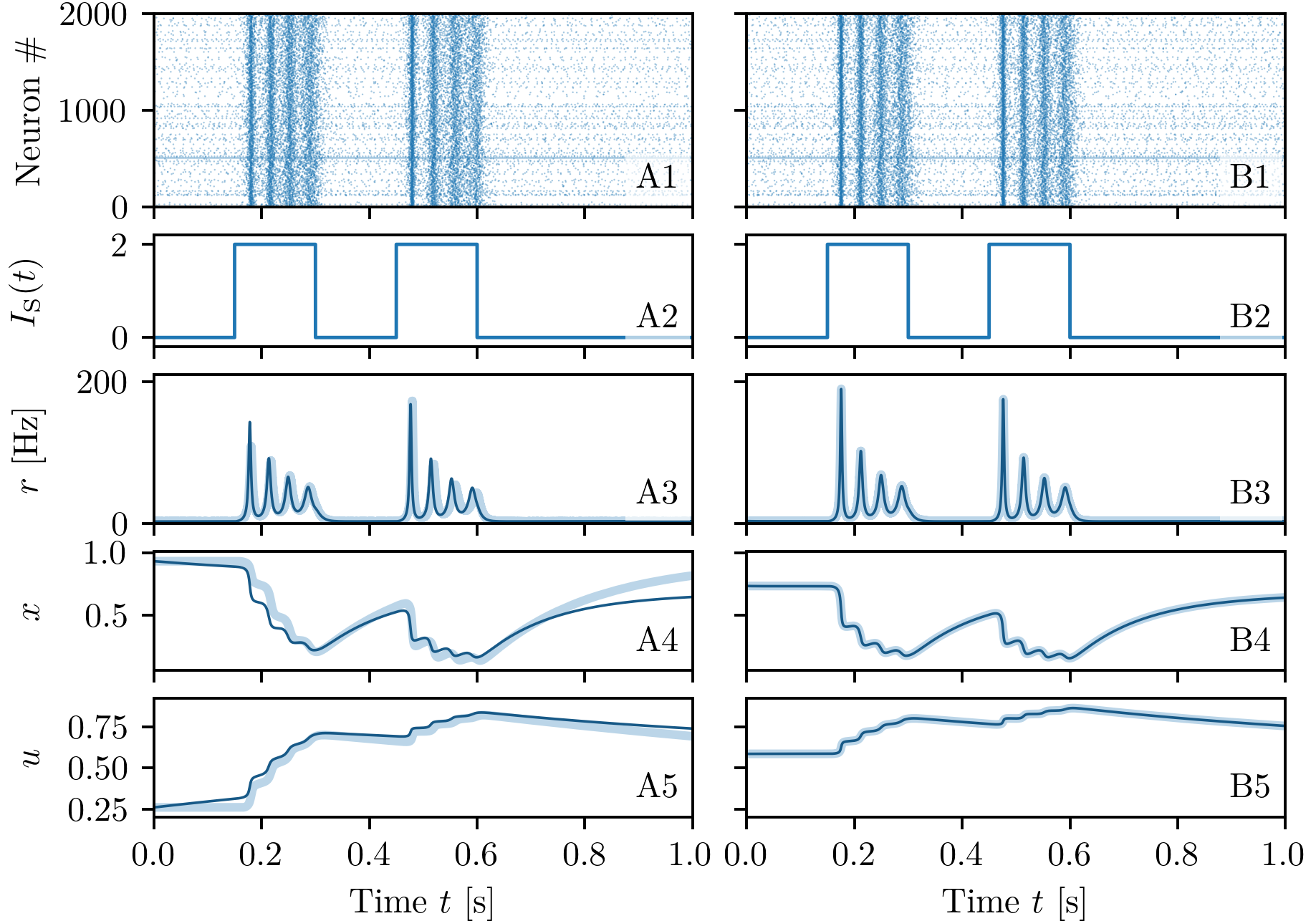}
\caption{{\bf Comparison among neural mass and network models}
The results of the neural mass model (solid line)  are compared with the network simulations
(shaded lines) for a single excitatory population with $\upmu$-STP  (column A) and with m-STP (column B). 
The corresponding raster plots for a subset of 2000 neurons are reported in ({\bf A1-B1}), 
 the profiles of the stimulation current $\Istim(t)$ in ({\bf A2-B2}), the instantaneous population
firing rates $r(t)$ in ({\bf A3-B3}), the available synaptic resources $x(t)$ in ({\bf A4-B4}),
and the utilization factors $u(t)$ in ({\bf A5-B5}). Simulation parameters are \mbox{$\taum^e=15$~ms}, \mbox{$H=0$}, \mbox{$\Delta=0.25$}, \mbox{$J=15$}, \mbox{$\Iback=-1$} and network size \mbox{$N=\mathrm{200,000}$}.
}
	\label{fig:comparison1}
\end{figure}

The neural mass model \eqref{eq:fre} reproduces exactly the population dynamics of a network of QIF spiking neurons, \red{in absence of STP},
in the infinite size limit, as analytically demonstrated and numerically verified in \cite{montbrio2015macroscopic}.
In order to verify if this statement is valid also
in presence of STP, we will compare for a single excitatory neural population 
the macroscopic evolution given by Eqs. \eqref{eq:fre} and \eqref{eq:dumeanfied}
with those obtained by considering QIF networks  with synaptic plasticity implemented at a
microscopic ($\upmu$-STP) and at a mesoscopic (m-STP) level. The results of these comparisons
are reported in Fig. \ref{fig:comparison1}: column A (B) for $\upmu$-STP (m-STP).

A spiking network of $N$ neurons with $\upmu$-STP is described by the evolution of
$3 N$ variables, since each neuron is described in terms of its membrane 
voltage and the two synaptic variables $U_i(t)$ and $X_i(t)$ accounting for the dynamics 
of its $N$ efferent synapses. The explicit dynamical model is reported in 
 Eqs. \eqref{eq:qifnetwork1}, \eqref{eq:du1} and \eqref{eq:splastic} in  sub-section \nameref{sec:stp}
 in  \nameref{sec:methods}.
In the m-STP model, the dynamics of all synapses is treated as a mesoscopic variable
in terms of only two macroscopic variables, namely $u(t)$ and $x(t)$. In this case 
the network model reduces to 
a set of $N+2$ ODEs, namely Eqs. \eqref{eq:qifnetwork1} and \eqref{eq:seminetwork} reported
in sub-section \nameref{sec:stp} in \nameref{sec:methods}.

In order to compare the different models we examine their response to the same delivered stimulus $\Istim(t)$,
which consists of two identical rectangular pulses of height $\Istim = 2$ and duration $0.15$ s separated by a time interval
of $0.15$ s (see Fig. \ref{fig:comparison1} (A2-B2)). The stimulus is presented when the system is in a quiescent state, i.e. it is characterized by a low firing rate. For both comparisons the neural mass model has been integrated starting from initial values of $r$,$v$,$x$ and $u$ as obtained from the microscopic state of the considered networks 
(for more explanations see sub-section \nameref{sec:nm} in \nameref{sec:methods}). The response is the same in both network models: each pulse triggers a series of four PBs of decreasing
amplitude followed by low activity phases in absence of stimuli (see Fig. \ref{fig:comparison1} (A1-B1)).
As shown in Fig. \ref{fig:comparison1} (A3-B3), the population firing rate $r(t)$ obtained from the neural mass model (solid line) is  almost coincident with the results of  the network simulations both with $\upmu$-STP and m-STP (shaded curves).
For all models, the average synaptic variables $x$ and $u$ displayed in Fig. \ref{fig:comparison1} (A4-B4) and (A5-B5) reveal
the typical temporal evolution of STP upon excitatory stimulation: the available resources $x$ (utilization factor $u$) decrease (increases) due to the series of emitted PBs in a step-wise fashion. Furthermore, in the absence of stimulation $x$ and $u$ tend to recover to
their stationary values $x=1$ and $u=U_0$ for $\upmu$-STP ($x=0.73$ and $u=0.59$ for m-STP) over time-scales dictated by $\taud$ and $\tauf$, respectively. Due to the fact that
$\taud \ll \tauf$, the synapses remain facilitated for a time interval $\simeq 1$ s after the pre-synaptic resources have recovered
from depression. The time courses of $x$ and $u$ for the neural mass compared to the network ones, shown in Fig. \ref{fig:comparison1} (A4-B4) and (A5-B5),
reveal an almost perfect agreement with the simulations of the network with m-STP, while some small discrepancies are observable 
when compared with the network with $\upmu$-STP. \red{We have verified that these discrepancies are not due to
finite size effects, therefore by further increasing the network size, the agreement will not improve. Instead,
these residual discrepancies are probably due to} the correlations and fluctuations of the microscopic variables $\{X_i\}$ and $\{U_i\}$, which are not taken into account in the mesoscopic model for STP \cite{tsodyks1998neural}, as pointed out in \cite{schmutz2020} (see also the discussion on this aspect reported in sub-section \nameref{sec:stp} in \nameref{sec:methods}).
We can conclude this sub-section by affirming that the developed neural mass model reproduces in detail the macroscopic dynamics of spiking networks of QIF neurons with both $\upmu$-STP and m-STP, therefore in the following we can safely rely on the mean-field simulations to explore the synaptic mechanisms at the basis of WM.

%\clearpage
\subsection*{Multi-Item Architecture}
\label{sec:ark}

In order to illustrate the architecture employed to store  more than one item simultaneously in WM, we will analyse the simplest non-trivial situation where we want to store two items.
For this purpose, we consider two excitatory populations with STP employed to store one item each,
and an inhibitory one, necessary to allow for item competition and a homoeostatic regulation of the firing activity.
In particular, we restrict to non-overlapping memories: neurons belonging to a certain excitatory population 
encode only one working memory item.  Furthermore, incoming information on a WM item, in form of an external stimulus, 
target only the excitatory population which codes for that item, hence making the response selective.

\begin{figure}[!h]
\includegraphics[width=1\linewidth]{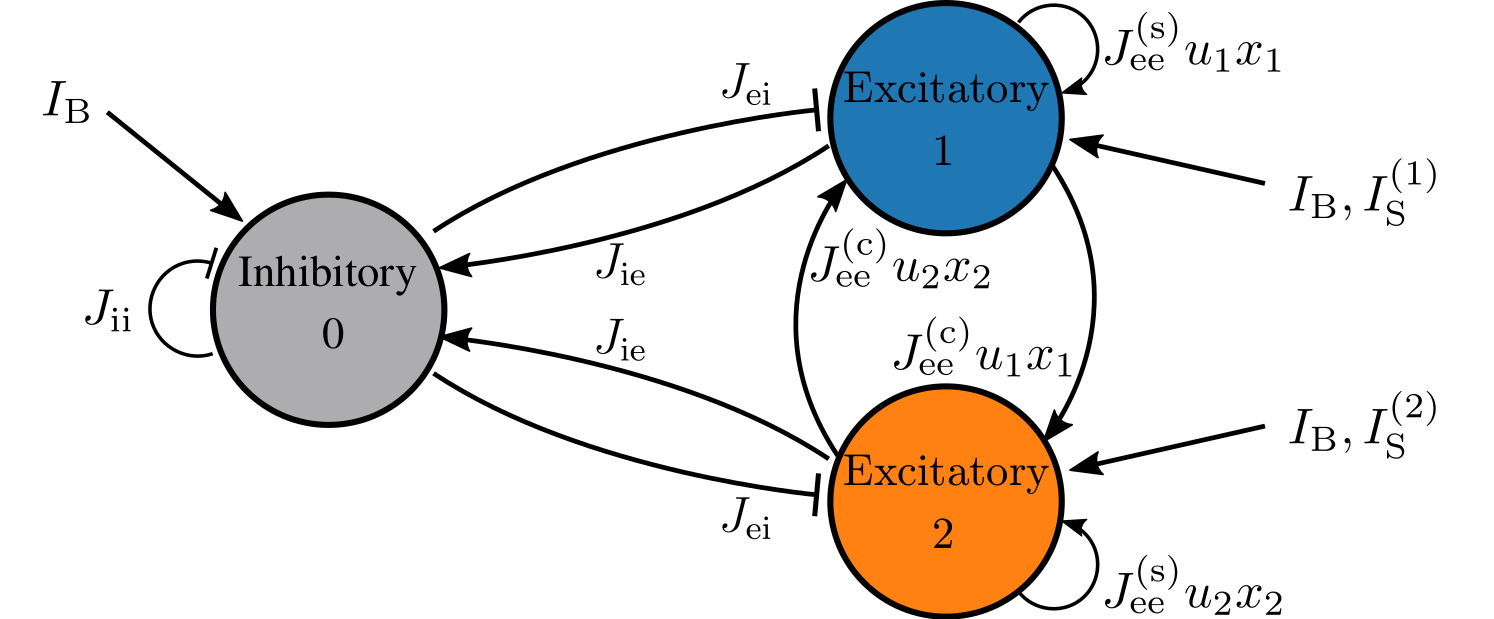}
\caption{{\bf Two-item architecture} The reported network is composed of two identical and \red{mutually} coupled excitatory populations and an inhibitory one: this architecture can at most store two WM items, one for each excitatory population.}
\label{fig:topology}
\end{figure}

As previously mentioned, only excitatory-excitatory synapses are plastic, therefore we will have
no time dependence for synaptic strength involving the inhibitory population.
Moreover, the WM items should be free to compete among them on the same basis, therefore we assume that the synaptic couplings
within a population of a certain type (inhibitory or excitatory) and among populations of the same type are identical.
In summary, we have 
\begin{eqnarray}
&& {\tilde J}_{00}(t)=J_{ii} , \enskip {\tilde J}_{0k}(t) = \Jie  , \enskip {\tilde J}_{k0}(t) = \Jei 
   , \enskip {\rm for} \enskip k > 0 \quad ; \\
&& {\tilde J}_{kk}(t)=\Jeep x_k(t) u_k(t) , \enskip  J_{kj}(t) = \Jeeb x_j(t) u_j(t)  , \enskip 
  \enskip   \forall k,j >0
  \nonumber
\end{eqnarray}
where we have denoted inhibitory (excitatory) populations by i (e) index.
Furthermore, synaptic connections within each excitatory population, coding for a certain memory, will be stronger than 
connections between different excitatory populations, thus we assume $\Jeep > \Jeeb$.
A common background input current $\Iback$ impinges on all three populations, thus playing a crucial role in controlling 
the operational point of the network. The selective storage and retrieval of memory items induced by incoming stimuli
is mimicked via time dependent external currents $\Istim^{(k)}(t)$ applied to the excitatory populations. 
The discussed multi-item architecture is displayed in  Fig. \ref{fig:topology}.

Due to the fact that each excitatory population is indistinguishable from the others and we have
only one inhibitory pool, we set for clarity $H_0 = H^{(\mathrm{i})}$, $H_k = H^{(\mathrm{e})} \enskip \forall k >0$ 
and $\Delta_0 = \Delta^{(\mathrm{i})}$, $\Delta_k = \Delta^{(\mathrm{e})} \enskip \forall k >0$. Furthermore, in order 
to better highlight the variations of the synaptic parameters, in the following we report normalized available resources ${\tilde x}_k$ and normalized facilitation factors ${\tilde u}_k$ \footnote{These variables are normalized with respect to the difference of their maximal and minimal value taken within the displayed time interval.}.

\subsection*{Memory maintenance with synaptic facilitation}
\label{sec:maint}

\begin{figure}[htp]
	\begin{adjustwidth}{-2.25in}{0in}
		\includegraphics[width=1\linewidth]{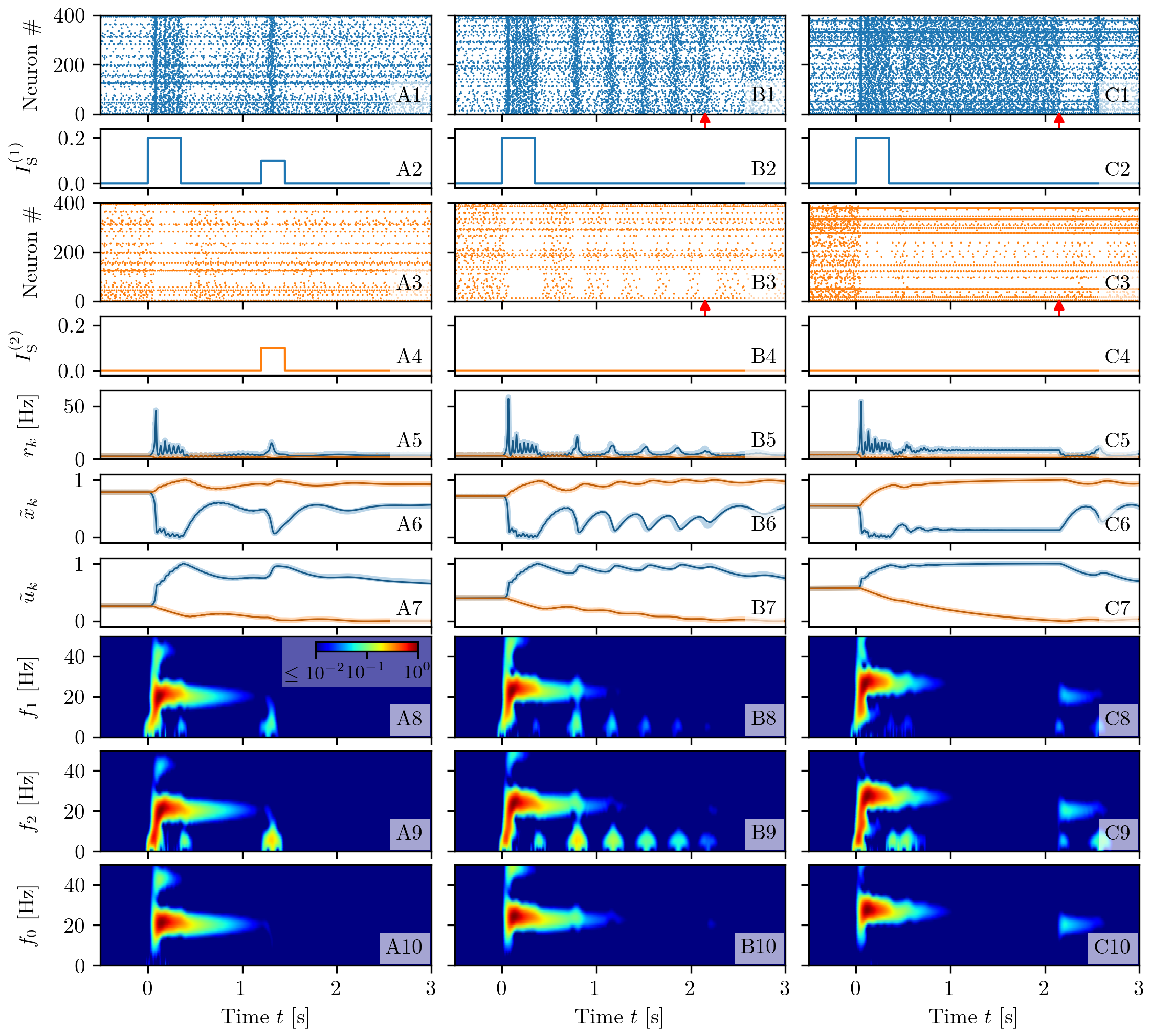}
\caption{{\bf Memory loading, maintenance and rehearsal} The results of three experiments are reported here for different background currents: selective reactivation of the target population ($\Iback=1.2$, A); WM maintenance via spontaneous reactivation of the target population ($\Iback=1.532$, B) and via a persistent asynchronous activity 
($\Iback=2$, C). Raster plots of the network activity for the first (blue, {\bf A1-C1}) and second (orange, {\bf A3-C3}) excitatory population; \red{here the activity of only 400 neurons over 200,000 ones is shown for each population}. 
Profiles of the stimulation current $\Istim^{(k)}(t)$ for the first  ({\bf A2-C2})
(second ({\bf A4-C4})) excitatory population.  Population firing rates $r_k$(t) ({\bf A5-C5}), normalized available resources 
${\tilde x}_k(t)$ ({\bf A6-C6}) and normalized utilization factors ${\tilde u}_k(t)$ ({\bf A7-C7}) of the excitatory populations calculated from the simulations of the neural mass model (solid line) and of the network (shading). Spectrograms of the mean membrane potentials $v_1(t)$ ({\bf A8-C8}), $v_2(t)$ ({\bf A9-C9}), and $v_0(t)$ ({\bf A10-C10}) obtained from the neural mass model; \red{for clarity the frequencies in these three cases have been denoted as $f_1$,$f_2$ and $f_0$, respectively.}  Red arrows in columns (B) and (C) indicate the time $t=2.15$ s at
which the background current is set to the value $\Iback=1.2$ employed in column (A). \red{The network
simulations have been obtained by considering three populations of \mbox{$N=\mathrm{200,000}$} neurons each (for a total of 600,000 neurons) arranged with the architecture displayed in Fig. \ref{fig:topology}}. Other parameters: \mbox{$H^{(i)}=H^{(e)}=0$}, \mbox{$\Delta^{(i)}=\Delta^{(e)}=0.1$}, \mbox{$\Jeeb=5 \sqrt{a}$}, \mbox{$\Jeep=35 \sqrt{a}$}, \mbox{$\Jie=13 \sqrt{a}$}, \mbox{$\Jei=-16 \sqrt{a}$}, \mbox{$\Jii=-14 \sqrt{a}$} with $a=0.4$. 
}
\label{fig0}
\end{adjustwidth}
\end{figure}

To show the capability of our model to reproduce WM activity we report in this sub-section 
three different numerical experiments. In particular, we will show that the model is
able to reproduce specific computational operations required by WM: namely, 
memory load, memory maintenance during a delay period, selective 
or spontaneous reactivation of the memory and memory clearance.
\red{The results of this analysis are reported in Fig. \ref{fig0}, where we compare
the neural mass simulations with the network dynamics with m-STP
for the topological architecture displayed in Fig. \ref{fig:topology}.
More specifically, for the network simulations we considered three coupled populations 
for a total of 600,000 neurons.
An analogous comparison is reported also in Fig.  \ref{fig:two_item_read_out},
where the competition between two memory items is studied.
}

\red{
\subsubsection*{Selective Reactivation}
}

Let us start from an example of selective reactivation of the target item via a non specific
read-out signal, targeting both excitatory populations, as shown in the left column (A) of Fig. \ref{fig0}.
Since the observed dynamics does not depend on the population into which an item is loaded, 
without loss of generality we always load the item in population one (coded by the blue colour in Fig.  \ref{fig0}).
The system is initialized in an asynchronous state of low firing activity common to both populations (as shown in panel (A5)),
therefore the synaptic variables have equal values (namely, $\tilde{x}_1 = \tilde{x}_2 = 0.79$ and $\tilde{u}_1=\tilde{u}_2=0.26$) dictated by the average firing rate (see panels (A6-A7)).

The memory load is performed at time $t=0$ s, by delivering a stimulation current in the form of a step with amplitude $\Delta I^{(1)}=0.2$ and width $\Delta T_1=350$ ms only to population one (panel (A2)). 
In response to this stimulus, the population displays PBs, as shown by the raster plot in panel (A1). The frequency of these COs is in the $\beta$ band
(namely, $\simeq 21.6$ Hz), as evident from the spectrogram in panel (A8). 
\red{In addition to this, the PBs in population one trigger PBs in the inhibitory population with a delay of $\simeq 1-2$ ms,
while the same COs' frequency is maintained, as evident from the inhibitory population spectrogram in panel (A10).
Therefore, the PBs are generated via a mechanism analogous to the pyramidal
interneuronal network gamma (PING) one \cite{tiesinga2009}, despite the fact that the frequency of the COs is now in the $\beta$ band.}

The intense firing of neurons in population one changes the internal state of its efferent synapses, thus leading to an initial drop (increase) of $x_1$ ($u_1$) (panels (A6-A7)).
Depression prevails on short time scales $\simeq \taud$, while at longer times $t > 0.5$ s, $\tilde{x}_1$ recovers to \red{almost} its initial
value and the synapses remain facilitated, with $\tilde{u}_1 > 0.73$, for one or two seconds (see panel (A7)).

Population two (denoted by the orange colour in  Fig.  \ref{fig0}) is not particularly affected by
the memory loading. Indeed, population two shows only a slight reduction in its asynchronous activity, 
which is reflected in a small increase (decrease) of $\tilde{x}_2$ ($\tilde{u}_2$) (see panels (A3) and (A6-A7)).
This is due to the action of the inhibitory bursts, which are unable to support PBs in population two, 
but sufficient to modulate its asynchronous activity as revealed by the spectrogram in panel (A9).

%These PBs on one side are unable to support bursts also in population two,
%for the unsufficient external stimulation, but on another side modulates its asynchronous firing, as revealed by the spectrogram in panel (A9).

Since the information on the initial stimulus is stored in the facilitation of the synapses, we expect
that, by presenting a weak non-specific read-out signal to both populations,
the memory will be reactivated, even if the population activity is back to the spontaneous level.
Indeed, if after a delay period of 1.2 s from the initial memory load, we inject a weak 
stimulation current of amplitude 0.1 for a time interval $\Delta T = 250$ ms to all the
excitatory neurons, we observe that only neurons of the blue population 
respond by emitting a PB of brief duration. The other neurons not associated to the loaded item 
remain at the baseline activity level, despite the stimulus (panel (A3)). 
The emission of the PB in population one has also the effect to refresh the memory,
i.e. the utilization factor $u_1$ which has decreased towards the initial
value during the delay period, returns to the value reached when the item was loaded. 
Therefore the memory can be maintained for a longer time period.

\red{
\subsubsection*{Spontaneous Reactivation}
}
 
In a second example, shown in the central column (B) of Fig. \ref{fig0},
we consider a case where the target population reactivates
spontaneously by emitting a regular series of PBs even in absence of read-out 
signals. To obtain such situation we increase the background signal \red{from $\Iback=1.2$} up to $\Iback=1.532$,
thus the system is in a regime where the spontaneous low activity state coexists with states in which one of the excitatory populations periodically emits PBs (for more details on the emerging states see the sub-section \nameref{sec:bif} within the section \nameref{sec:methods}).
As in the previous experiment, the system is initialized in a low firing activity state, 
where only spontaneous activity is present. The memory item is loaded in population one (see panels (B1-B5)).
In this case population one encodes the item by emitting a series of PBs with a slightly higher
frequency ($\simeq 24.1$ Hz) with respect to the previous case, due to the increase in $\Iback$, but still
in the $\beta$ range (as shown in panels (B8-B10)). After a short time delay, PBs emerge regularly in a self-sustained manner: each reactivation leads 
to an increase of $\tilde{u}_1$ and a decrease of $\tilde{x}_1$. The time interval between two PBs is
dictated by the time required to recover sufficient synaptic resources in order to emit a new PB,
i.e. by the time scale $\taud$ controlling the depression \cite{tsodyks2000,luccioli2014} 
(as shown in panel (B6)). Thus, we have  memory maintenance through synaptic facilitation, which is refreshed by PBs.
As one can appreciate from the spectrogram of population one reported in panel (B8),
whenever a PB is delivered, a transient oscillation in the $\delta$ band,
at a frequency $\simeq 3$ Hz, is observed. Similar transient oscillations have been observed
when items are loaded in the PFC of monkeys during WM tasks \cite{siegel2009phase}.
Signatures of these oscillations are present also in the spectrogram of the other excitatory population (panel (B9)) but,
in this case, they are due to a modulation of the subthreshold membrane potentials and not to PBs.
It is interesting to note that $\delta$-oscillations are only sustained by the activity of the excitatory
populations, since in the inhibitory spectrogram (panel (B10)) there is no trace of them.

The sequence of PBs can be terminated by reducing the excitation $\Iback$: this operation is performed at 
time $t=2.15$ s after the initial stimulation and signalled by the red arrow in panels (B1) and (B3). 
Also in this experiment the memory load and retrieval is selective, since the second population activity
is almost unaffected by these operations as it turns out from the raster plot reported in panel (B3).

\red{
\subsubsection*{Persistent Activity}
}

As a third experiment, we consider a situation where the memory item is
maintained via the persistent activity of the target population, as 
shown in the right column (C) of Fig. \ref{fig0}.
This is possible when the non-specific background input $\Iback$ is increased further
such that we can observe multistability among three asynchronous states: one corresponding
to the spontaneous activity at low firing rate of both excitatory populations and
the other two corresponding to population one (two) in a persistent state with increased firing activity 
and population two (one) at a spontaneous level of activity. More details can be found in \nameref{sec:bif}.
When at time $t=0$ the memory item is loaded in population one, the population responds
by emitting a series of PBs at a frequency $\simeq 27.2$ Hz in the $\beta$-$\gamma$ band
(see panels (C8-C10)).
As shown in panels (C1) and (C5), once the loading is terminated the blue population enters into a regime of persistent firing characterized by an almost constant firing rate $r_1 \simeq 8.6$ Hz joined to constant values of the synaptic variables. In particular, the available resources of population one
remain at a constant and low value due to the continuous firing of the neurons ($\tilde{x}_1 = 0.13$) (see panel C6).
However, the lack of resources is compensated by the quite high utilization factor $\tilde{u}_1=0.98$
(reported in panel (C7)). Indeed in this case the memory is continuously refreshed by the persistent spiking activity. 
At time $t=2.15$ s, indicated by red arrows in panels (C1) and (C3), the value of $\Iback$ is reduced to $\Iback=1.2$ and the persistent activity is interrupted. This would correspond to a memory clearance if WM would be based on the spiking activity only. Instead in this case, it represents only the interruption of the persistent activity, the memory clearance occurring 
when the facilitation returns to its original value after few more seconds (the decay of $u_1$
is evident in panel (C7)). The second population, not involved in the memory loading, remains always in a low firing rate regime
(panels (C3) and (C5)).
 
 As expected, the neural mass dynamics reproduces almost perfectly the network dynamics with m-STP for all the three considered experiments. In particular, this can be appreciated through the agreement among shaded lines, corresponding to
network simulations, and solid ones, referring to the neural mass evolution,
reported  in Fig. \ref{fig0} (A5-C5),(A6-C6), (A7-C7). 

\red{Furthermore, it should be remarked that memory loading is characterized 
by similar spectral features for all the three experiments. As shown in Figs. \ref{fig0} (A8-B8-C8) and (A9-B9-C9),
a transient broad-band response of the excitatory populations is observable in the range 3-18 Hz,
locked with the stimulation onset and followed by a steady-state activity in the $\beta$-$\gamma$ range
(namely 21-27 Hz), which lasts for the whole duration of the stimulation. The broad-band oscillations emerge 
due to the excitation of the harmonics of a fundamental frequency $\simeq 2-3$ Hz, associated to 
one item loading in the memory. Similarly, the $\beta$-$\gamma$ activity is initiated by memory
loading that induces, in this case, damped oscillations towards a focus equilibrium state in the stimulated excitatory population: the damped oscillations are sustained by the inhibitory pool via a PING-like mechanism. Quite astonishingly, similar evoked power spectra have been reported for stimulus-locked EEG responses to vibrotactile stimuli in primary somatosensory 
cortex in humans \cite{spitzer2010}. In more details, as shown in Fig. 1B of Ref. \cite{spitzer2010},
the stimulus onset induces a broad-band activity in the 4-15 Hz range followed by a stationary activity 
at $\simeq 26$ Hz during the vibrotactile stimulations.
}

\red{
	\subsubsection*{Comparison with a heuristic firing rate model}  
	\label{comparison_fr}	
	To close this sub-section, let us compare the population dynamics obtained by employing the neural mass model with STP (Eqs. \eqref{eq:fre},\eqref{eq:dumeanfied}) and  a heuristic firing rate model developed to mimic the dynamics of a QIF network with m-STP (for more details see sub-section \nameref{sec:wilsoncowan}). Recent studies have shown that this firing rate model, in absence of plasticity, is unable to capture some macroscopic behaviour displayed by the corresponding QIF networks \cite{devalle2017firing,schmidt2018network}. In particular, it does not reproduce fast COs present in inhibitory networks \cite{devalle2017firing} and it does not feature memory clearance via nonlinear resonance with an external $\beta$-forcing, contrary to the spiking network under forcing \cite{schmidt2018network}. Here, we want to understand which network's dynamics are eventually lost by employing such a heuristic model with STP.}

\begin{figure}[htp]
	\includegraphics[width=1\linewidth]{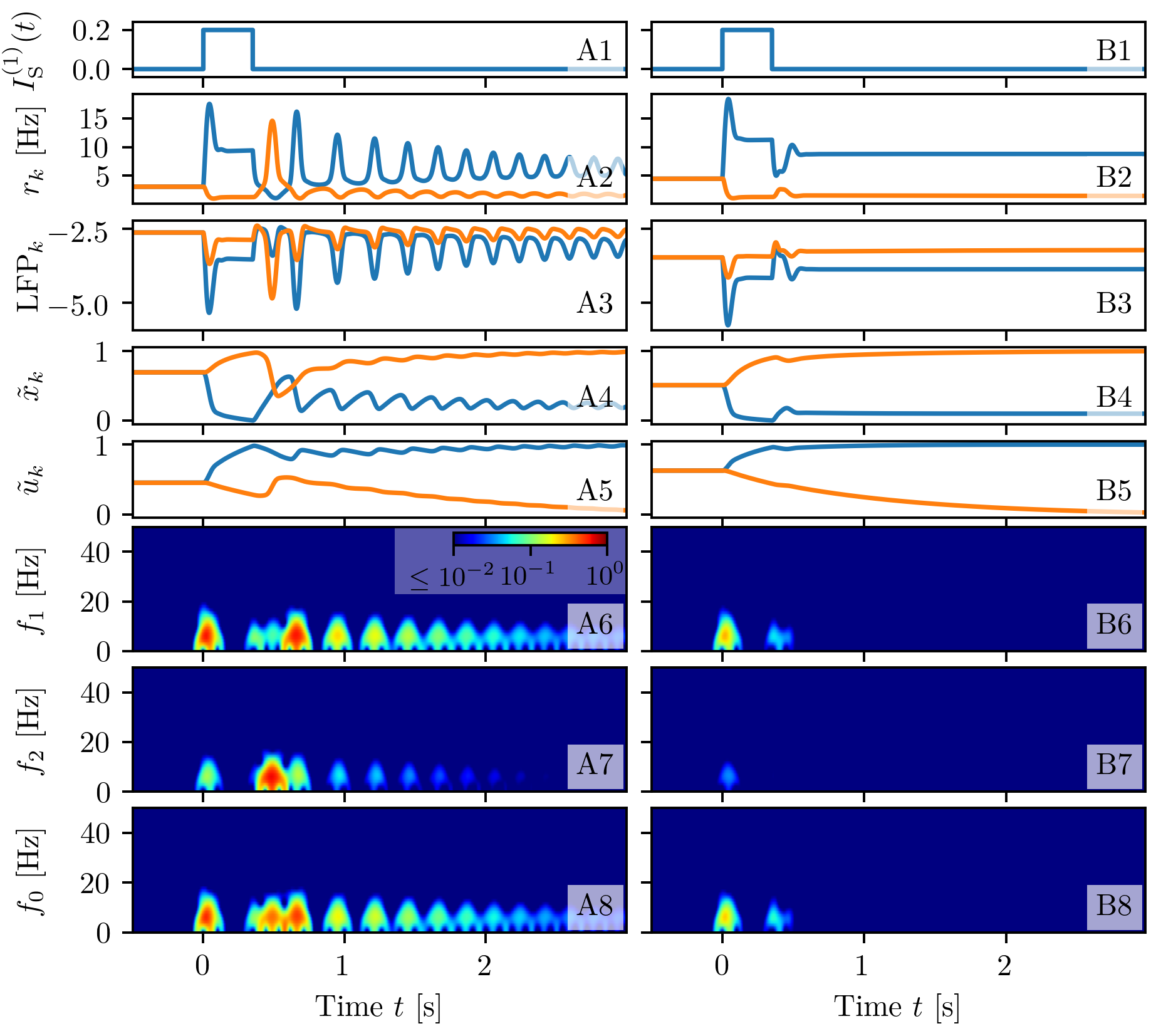}
	\caption{\red{{\bf WM operations for a heuristic firing rate model} 	
	The results of two experiments are reported here for different background currents: WM maintenance via spontaneous reactivation of the target population ($\Iback=1.520$, A) and via a persistent asynchronous activity 
	($\Iback=2.05$, B). Profiles of the stimulation current $\Istim^{(k)}(t)$ for the excitatory populations ({\bf A1-B1}). Population firing rates $r_k$(t) ({\bf A2-B2}), local field potentials $\mathrm{LFP}_k$ defined in Eqs. \eqref{LFP} ({\bf A3-B3}),  normalized available resources  ${\tilde x}_k(t)$ ({\bf A4-C4}) and normalized utilization factors ${\tilde u}_k(t)$ ({\bf A5-C6}) of the excitatory populations calculated from simulations of the firing rate model  \eqref{eq:wc1},\eqref{eq:sfi2}.
Spectrograms of the local field potentials:  $\mathrm{LFP}_1(t)$ ({\bf A6-B6}), $\mathrm{LFP}_2(t)$ ({\bf A7-B7}), and $\mathrm{LFP}_0(t)$ ({\bf A8-B8}). All the other parameter values as in Fig. \ref{fig0}.}}	 
	\label{fig:wilsoncowan}
\end{figure}

\red{ 
Therefore we have repeated the experiments leading to memory maintenance via spontaneous reactivation and
persistent activity, previously reported in Figs. \ref{fig0} (B) and (C), with the same topological configuration.
The results of this analysis are shown in Fig. \ref{fig:wilsoncowan}: namely, Column (A) is devoted to 
spontaneous reactivation and Column (B) to persistent activity. In more detail, the multi-population network is initialized in the quiescent state with asynchronous activity and low firing rates (see panels (A2-B2)); at time \mbox{$t=0$ s}, a 
current step  of amplitude \mbox{$\Delta I^{(1)}=0.2$} and time width \mbox{$\Delta T_1=350$ ms} is injected into population one (blue line), as shown in panels  (A1-B1).
When a background current $\Iback=1.520$ is chosen, as soon as the stimulation is removed, population one enters into a cycle of periodic PBs, each one refreshing the synaptic facilitation and allowing for the memory maintenance (column (A)).
By increasing the current to $\Iback=2.05$,  the stimulus leads population one into a state of persistent firing activity, which maintains the synaptic variables at almost constant values (column (B)). Therefore, the heuristic firing rate model is able to reproduce the WM operations performed by the neural mass model.
However, when comparing the firing rate time traces of the QIF network and of the neural mass model (reported in Figs. \ref{fig0} (B5-C5)) with the ones corresponding to the heuristic firing rate model (shown in Figs. \ref{fig:wilsoncowan} (A2-B2)), we note that, after the onset of stimulation, the population firing rates of the QIF network and of the neural mass model exhibit fast damped COs in the $\beta$-$\gamma$ range, which are completely absent in the heuristic rate model.
These findings are confirmed by the spectrograms\footnote{\red{The spectrograms for the heuristic model have been evaluated for the 
local field potentials $\mathrm{LFP}_k$ defined in Eqs. \eqref{LFP} and reported in Fig. \ref{fig:wilsoncowan} (A3-B3);
this is due to the fact that the rate model cannot provide any information on the membrane potentials.}
} (compare Figs. \ref{fig0}(B8-B10) and (C8-C10)
with Figs. \ref{fig:wilsoncowan} (A6-A8) and (B6-B8)), which reveal a
a vanishingly small power in the $\beta$-$\gamma$ bands for the heuristic model in both experiments.
}

\red{The reason for this absence is related to the fact that the stimulation leads population one in
an excited state, which turns out to be a stable node equilibrium and not a focus, as for the neural mass model; therefore
in this case it is not possible to observe transient PBs associated to memory loading. 
As a consequence, the heuristic model does not show any activity in the $\beta$ and $\gamma$ bands,
despite this kind of activity has been reported experimentally in the PFC of monkeys
performing WM tasks \cite{lundqvist2016gamma} and in the primary somatosensory cortex of humans
due to vibrotactile stimuli \cite{spitzer2010}, and associated to memory loading and
recall in WM models \cite{dipoppa2013flexible}. Furthermore, as we will discuss in detail in the sub-section \nameref{7items}, the loading of many items in WM is usually associated with $\gamma$-power enhancement, as shown by various experiments \cite{howard2003gamma,van2010hippocampal,roux2012gamma}. While this effect
is present in our neural mass model, where the high frequency oscillations are enhanced due to a resonant mechanism with the transient oscillations towards the focus equilibrium, this aspect cannot be clearly reproduced by this heuristic firing rate model. 
}

\subsection*{Competition between two memory items}

\label{sec:competitiontwoitem}

In this sub-section we verify the robustness of the investigated set-up when two memory items are loaded,
one for each excitatory population. In particular, we will examine the possible outcomes of the
competition between two loaded items with emphasis on the mechanisms leading to memory juggling \cite{lewis2014}.
\red{More specifically, we will consider three different operational modes,
where the items are maintained in the WM due to different
mechanisms: namely, in the first case thanks to periodic stimulations; 
in the second one due to self-sustained periodic PBs and in the last one
to persistent spiking activity.}

\red{
\subsubsection*{Periodic Stimulations}
}

\begin{figure}[!h]
	\begin{adjustwidth}{-2.25in}{0in}
		\includegraphics[width=1\linewidth]{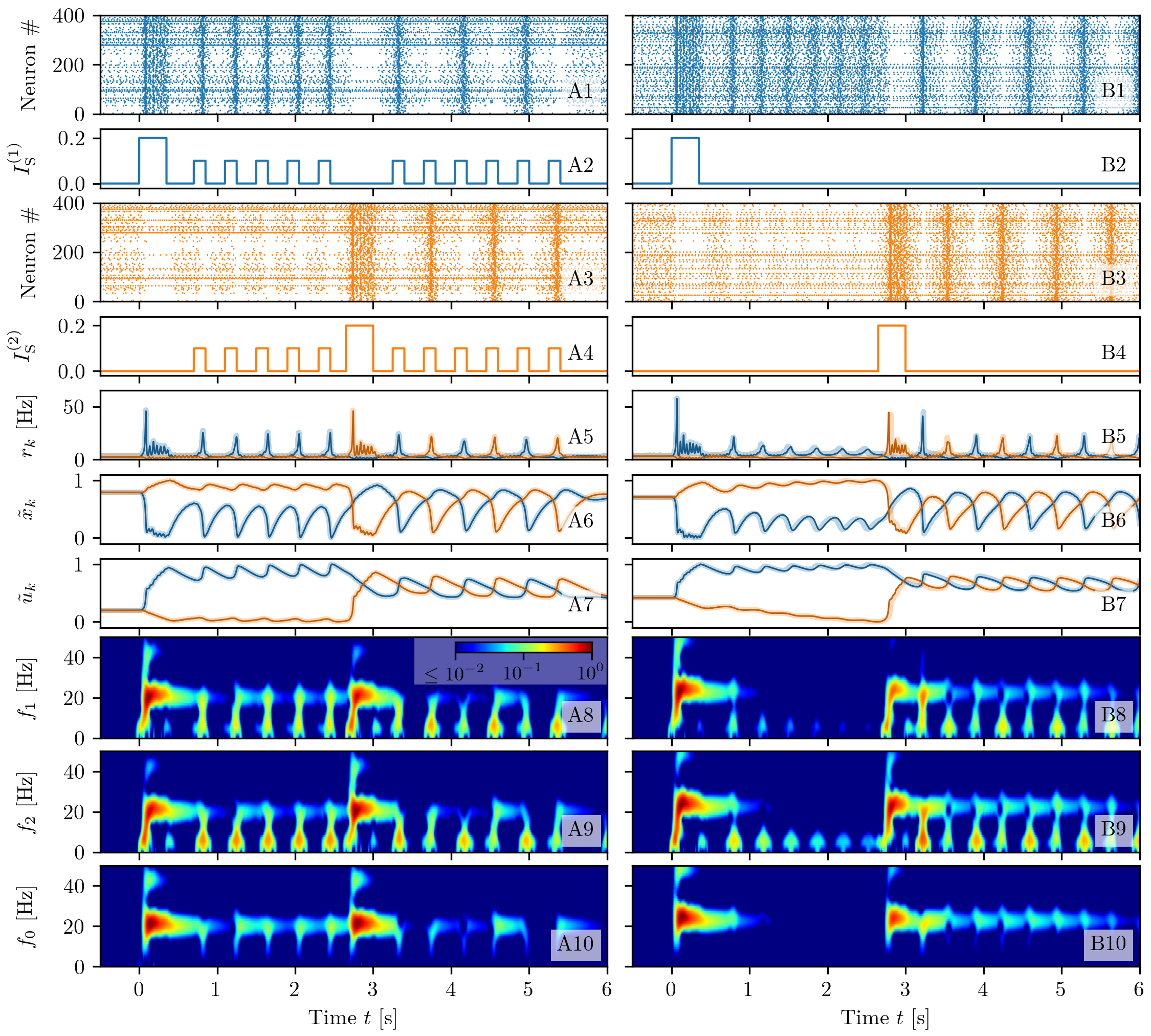}
		\caption{{\bf Juggling between two memory items} The memory juggling is obtained in two experiments with different background currents: in presence of a periodic unspecific stimulation ($\Iback=1.2$, A) and in the case of spontaneous WM reactivation ($\Iback=1.532$, B). Raster plots of the network activity for the first (blue, {\bf A1-B1}) and second (orange, {\bf A3-B3}) excitatory population. Profiles of the stimulation current $\Istim^{(k)}(t)$ for the first  ({\bf A2-B2})
(second ({\bf A4-B4}))) excitatory population.  Population firing rates $r_k$(t) ({\bf A5-B5}), normalized available resources 
${\tilde x}_k(t)$ ({\bf A6-B6}) and normalized utilization factors ${\tilde u}_k(t)$ ({\bf A7-B7}) of the excitatory populations calculated from the simulations of the neural mass model (solid line) and of the network (shading). Spectrograms of the mean membrane potentials $v_1(t)$ ({\bf A8-B8}), $v_2(t)$ ({\bf A9-B9}), and $v_0(t)$ ({\bf A10-B10}) obtained
from the neural mass model. All the other parameter values as in Fig. \ref{fig0}.}
		\label{fig:two_item_read_out}
	\end{adjustwidth}
\end{figure}

First we analyse the two-item memory juggling in presence of a periodic unspecific stimulation to the
excitatory populations. As shown in the left column (A) of Fig. \ref{fig:two_item_read_out}, 
at time $t=0$ we load the first item in population one
by stimulating the population for $\Delta T_1 = 350$ ms, with an excitatory step of amplitude $\Delta I^{(1)}=0.2$ (see panel (A2)).
Population one encodes the item via the facilitation of its efferent synapses. This is a 
consequence of a series of PBs emitted via \red{a PING-like mechanism} at a frequency $\simeq 21.6$ Hz in the $\beta$-range, as evident from 
the spectrograms in panels (A8-A10).
Subsequently a periodic sequence of unspecific stimulations of small amplitude and brief duration is delivered 
to both populations (namely, step currents of amplitude 0.1 and duration 150 ms applied at intervals of 400 ms). 
These are sufficient to refresh the memory associated to population one, that reacts each time, by emitting a brief PB
able to restore $\tilde{u}_1$ to a high value (panel (A7)). On the contrary population two remains in the low activity regime (panels (A3) and (A5)).  

The second item is loaded at time $t=2.65$ s, by presenting to population two a signal equal to the one presented
at time $t=0$ (panel (A4)). Also the loading of this item is associated to PBs emitted in the
$\beta$ range and involving the inhibitory population. During the presentation of the second item, the previous item is temporarily suppressed.
However, when the unspecific stimulations are presented again to both population, PBs are triggered in both populations, resulting to be
in anti-phase, i.e. PBs alternate between the two populations at each read-out pulse. The period related to each item is therefore twice the interval between read-out signals (see panels (A1) and (A3)).
This is due to the fact that a PB in one excitatory population stimulates the action of the inhibitory neurons in population zero which, as a consequence, suppresses the activity of the other excitatory population, leading to the observed juggling between the two items in working memory. As clear from the spectrograms in panels (A8-A10), the unspecific stimulations of the
excitatory populations induce localized peaks in their spectrograms around 2-3 Hz, not involving the inhibitory population. 
The latter instead oscillates at a frequency $\simeq 21.6$ Hz, thus inducing a modulation of the neural activity of the
two excitatory populations. This explain the presence of the series of peaks in the three spectrograms
around 20-25 Hz.

\begin{figure}[!h]
	\begin{adjustwidth}{-2.25in}{0in}
		\includegraphics[width=1\linewidth]{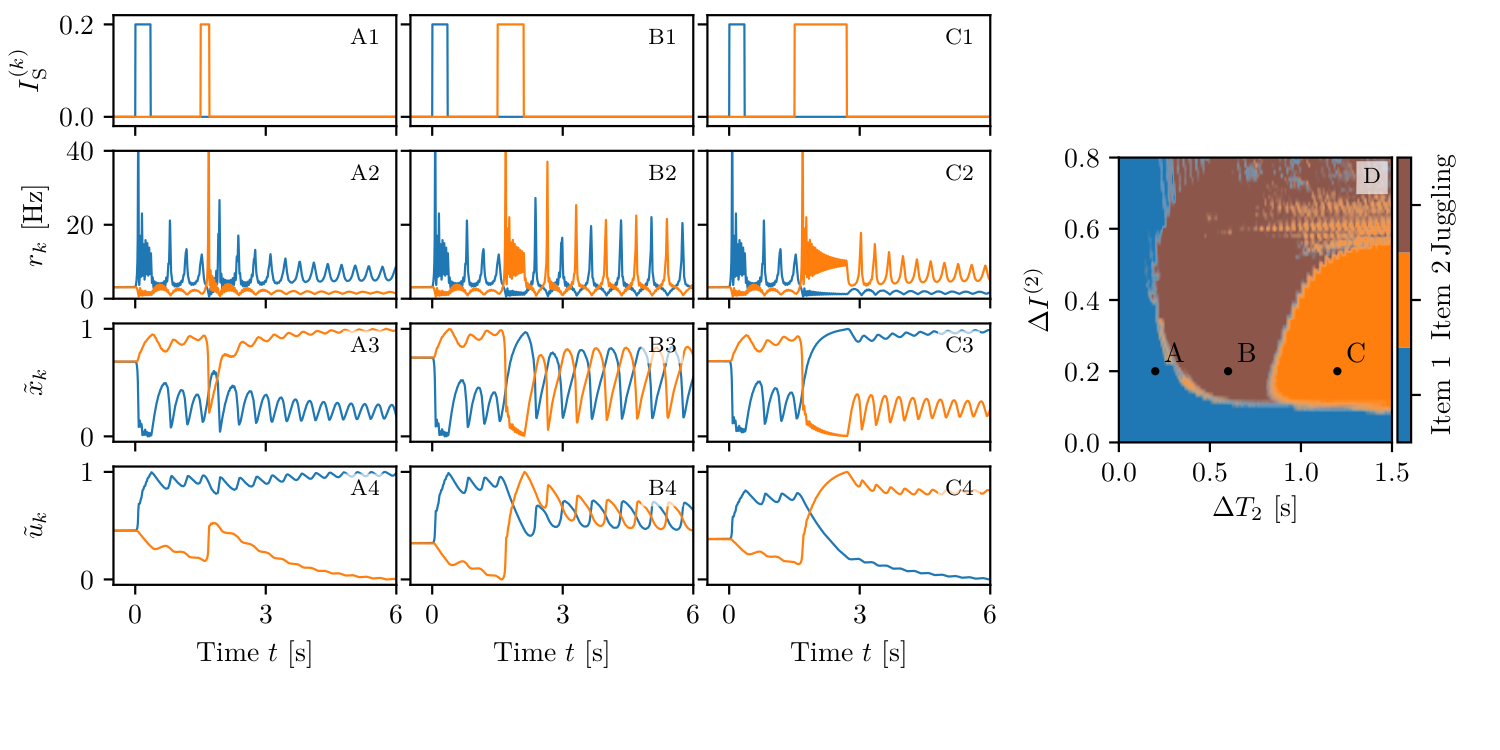}
		\caption{{\bf Competition between two memory items}. The loading of two memory items is performed in three different ways via two step currents of equal amplitudes $\Delta I^{(1)} = \Delta I^{(2)}= 0.2$, the first one delivered at $t=0$ and the second one at $t=1.5$ s. The first step has always a duration of $\Delta T_1=0.35$ s; the second one $\Delta T_2=0.2$ s (column A), $\Delta T_2=0.6$ s (column B) and $\Delta T_2=1.2$ s (column C). Stimulation currents $\Istim^{(k)}(t)$ ({\bf A1-C1}), instantaneous population firing rates $r_k(t)$ ({\bf A2-C2}), normalized available resources $\tilde{x}_k(t)$ ({\bf A3-C3}) and normalized utilization factors $\tilde{u}_k(t)$ ({\bf A4-C4}) for the first (blue) and second (orange) excitatory population. \red{Final memory states as a function of the amplitude $\Delta I^{(2)}$ and width $\Delta T_2$ of the stimulus to population two ({\bf D})}. In ({\bf D}) the three letters A, B, C refer to the states examined in the corresponding columns. The data reported in panel ({\bf D}) has been obtained by delivering the second stimulation at different phases $\phi$ with respect to the period of the CO of the first population: 20 equidistant phases in the interval  $\phi\in[0,2\pi)$ have been considered. For each of the three possible outcomes an image has been created with a level of transparency corresponding to the fraction of times it has been measured. The three images have been merged resulting in panel ({\bf D}). All the other parameter values as in Fig. \ref{fig0}, \red{apart for  $\Iback=1.532$}.}
\label{fig:distractor_juggling_switch}
\end{adjustwidth}
\end{figure}

\red{
\subsubsection*{Self-Sustained Population Bursts}
}

Let us now consider a higher background current (namely, $\Iback=1.532$), where self-sustained periodic PBs can be emitted by each excitatory population. In this case it is not necessary to deliver unspecific periodic stimuli
to refresh the synaptic memory. The loading of the two items is done by applying the same currents as before ($\Delta I^{(1)}=\Delta I^{(2)}=0.2$, $\Delta T_1=\Delta T_2=350$ ms) to the two populations, as shown in Fig. \ref{fig:two_item_read_out} (B2) and (B4). The memory loading occurs, once more, via series of bursts emitted with frequency in the $\beta$ range, through a 
\red{PING-like mechanism}. Once the item is loaded, a series of subsequent PBs is emitted regularly, at a frequency $\simeq 2.90$ Hz (see Fig. \ref{fig:two_item_read_out} (B1), (B5) and (B8-B10)).

The second population remains quiescent until the second item is loaded  (panels (B3) and (B4)).
During the loading period, the activity of the first population is temporarily suppressed while, at later times, both items are
maintained in the memory by subsequent periodic reactivations of the two populations. In particular it turns out that periodic reactivations are in anti-phase, with frequency $\simeq 1.5$ Hz, i.e., one observes juggling between the two working memory items (see panel (B5)). Also in this case, the inhibitory population is responsible for sustaining the $\beta$ rhythm, while the excitatory ones for the slow oscillations, as revealed by the spectrograms in panels (B8-B10).
\red{As testified by the comparisons reported in Fig.  \ref{fig:two_item_read_out} (A5-A7) and (B5-B7),
the agreement among the network simulations (solid lines) and the neural mass results (shaded lines) is impressive
even for the two experiments discussed sofar in this Section.}

For the setup analysed in column (B) of Fig. \ref{fig:two_item_read_out}, the competition between the two memory items can have
different final outcomes, depending on the characteristics of the stimuli, namely their amplitude and duration. For simplicity we fix these parameters for the first stimulation and vary those of the second stimulation. %depending on the characteristic of the second stimulus, namely its amplitude and time width. 
As shown in Fig. \ref{fig:distractor_juggling_switch} (A-C), three outcomes are possible for stimuli with the same amplitude $\Delta I^{(2)}=0.2$ but different pulse duration: item one wins, item two wins or they both coexist in the memory (juggling). If $\Delta T_2$ is too short, the facilitation $u_2$ has no time to become sufficiently large to compete with that of population one (column A). Therefore, once the second stimulus is removed, population one recovers its oscillatory activity and population two returns to the low firing asynchronous activity (see panel A2). For intermediate durations $\Delta T_2$, at the end of the stimulation period, the facilitation $u_2$ reaches the value $u_1$ (see panel B4), thus leading the two items to compete. Indeed, as shown in Fig. \ref{fig:distractor_juggling_switch} (B2), the two populations display COs of the same period but in phase opposition, analogously to the case reported in  Fig. \ref{fig:two_item_read_out} (B). The stimulation of the second population suppresses the activity of the first, leading to a relaxation of the facilitation to the baseline. Hence, whenever $\Delta T_2$ is sufficiently long, the facilitation of population two prevails on that of population one (see panel (C4)) and, as a final outcome,
population two displays COs, while population one has a low activity (panel (C2)).

The results of a detailed analysis for different amplitudes $\Delta I^{(2)}$ and durations $\Delta T_2$ are summarized in Fig. \ref{fig:distractor_juggling_switch} (D). Population one always receives the same stimulus at time $t=0$ ($\Delta I^{(1)} = 0.2$ and $\Delta T_1=0.35$ s), while $\Delta I^{(2)}$ and $\Delta T_2$  of the second stimulus, delivered at time $t=100$ s, are varied \footnote{The long interval of 100 seconds between the two stimuli is chosen in order to ensure that the outcome 
is independent of the characteristic of the first stimulus: namely, $\Delta I^{(1)}$ and $\Delta T_1$.}.
In particular, the amplitude has been varied in steps of 0.01 within the interval $\Delta I^{(2)}\in[0,0.8]$, while the duration in steps of 0.01 s for $\Delta T_2\in[0,1.5]$ s. The classification of the final states has been performed three seconds after the second stimulation, by estimating the time averaged firing rates $\left<r_1\right>$ and $\left<r_2\right>$ over an interval of one second. From this we get the indicator $P:=\frac{\left<r_{1}\right>}{\left<r_1\right>+\left<r_2\right>}$. If $P > 0.7$ ($P < 0.3$) item one (item two) has been memorized, otherwise we have juggling between the two items. From Fig. \ref{fig:distractor_juggling_switch} (D), we see that, for sufficiently small amplitudes $\Delta  I^{(2)} < 0.11$ or durations
$\Delta T_2 < 0.2$ s, the second stimulation is unable to change the memory state, that remains on item one. 
\red{This can be understood by looking at the phase diagram shown in Fig. \ref{bifurcation_diag} (B), from which it is clear that population two, being in the low firing regime, has a low value of the membrane potential $v_2$ and that such state is separated by the high activity regime from a barrier $\Delta v_2$, which is of the order of the distance from the saddle (dotted line in the figure).}
A frequent outcome of the experiments is the coexistence between the two memory items, that we observe in a large parameter region for intermediate pulse durations, namely 0.2 s $< \Delta T_2 < 0.8$ s. For sufficiently long perturbations $\Delta T_2 > 0.8$ s and intermediate amplitudes $ 0.11 < \Delta  I^{(2)} < 0.53$, item two is memorized. It is unexpected that, for larger amplitudes, we observe a multistability among all the three possible outcomes. Indeed, small variations of amplitude, duration or delivering time can induce a completely different outcome, thus suggesting that for these parameter values, some transient chaotic
behaviour is observable \cite{lai2011,cortes2013}.

\begin{figure}[!h]
	\begin{adjustwidth}{-2.25in}{0in}
	\includegraphics[width=1\linewidth]{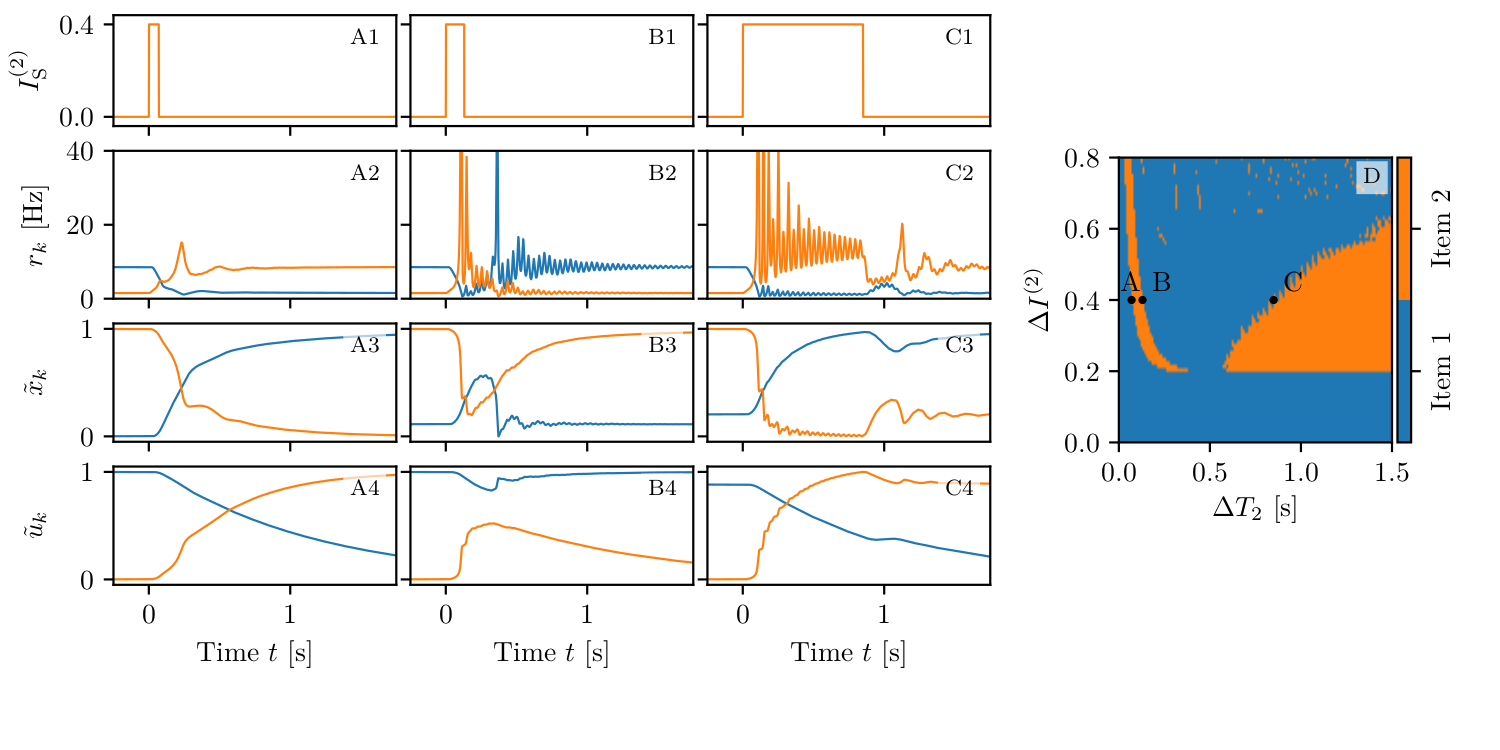}
	\caption{{\bf Memory item switching with persistent activity}. The three columns (A-C) refer to three values of $\Delta T_2$ for which we
	have a memory switching from one item to the other for a background current $\Iback=2$ supporting persistent state activity.
	(A) $\Delta T_2=70$ ms; (B) $\Delta T_2=130$ ms and (C) $\Delta T_2=850$ ms. 
	Profiles of the stimulation current $\Istim^{(2)}(t)$ for the second excitatory population  ({\bf A1-C1}). Population firing rates $r_l$(t) ({\bf A2-C2}), normalized available resources 	${\tilde x}_l(t)$ ({\bf A3-C3}) and normalised utilization factors ${\tilde u}_l(t)$ ({\bf A4-C4}) of the excitatory populations calculated from the simulations of the neural mass model. \red{Final memory states as a function of the amplitude $\Delta I^{(2)}$ and width $\Delta T_2$ of the stimulus to population two ({\bf D}). In ({\bf D}) the three letters A, B, C refer to the states examined in the corresponding columns.} Other parameters as in Fig. \ref{fig0}, \red{apart for $I_B=2$}.}
	\label{Sfig1}
	\end{adjustwidth}
\end{figure}

\red{
\subsubsection*{Persistent State Activity}
}

We have performed the same analysis for the higher current value $\Iback=2$, which supports a persistent state activity of one of the two populations. We set initially population one in the persistent state and we deliver a stimulus in form of a step current to population two. In this case we can obtain only two possible final outcomes: 
item one (two) loaded corresponding to population one (two) in the persistent state, while the other in the low activity regime.
No memory juggling has been observed with persistent states in our model. The results of this investigation are summarized in
Fig. \red{\ref{Sfig1} (D)}; for what concerns the final prevalence of item two, the results are similar to the ones
obtained in the previous analysis, where the item was encoded in self-sustained periodic COs. 
\red{The main differences with the previous case are: i) the minimal perturbation amplitude needed to observe an item switching, that is
now larger, namely $\Delta I^{(2)} \simeq 0.2$; ii) the existence of a narrow stripe in
the $(\Delta T_2, \Delta  I^{(2)})$-plane where item two can be finally selected for brief 
stimulation duration. The increased value of the minimal $\Delta I^{(2)}$ required to switch to item two is justified
by the fact that, for $\Iback=2$, the barrier $\Delta v_2$ that has to be overcome to access the high firing regime, is higher than before, as evident from the phase diagram reported in Fig. \ref{bifurcation_diag} (B). The origin of the stripe can be understood}
by examining the results reported in  Fig. \ref{Sfig1} \red{columns A-C}, for three stimulations with the same amplitude $\Delta I^{(2)} = 0.4$ and different durations. The effect of the stimulation is to increase the activity of population two and indirectly, also the inhibitory action on population one. However, for short $\Delta T_2 < 70$  ms, the activity of population two
is not sufficient to render its efferent synapses stronger than those of population one. As a matter of fact,
when the stimulation is over, population one returns into the persistent state. 
For longer $\Delta T_2 \ge 70$ ms, population one is not able to recover its resources before population two 
elicits a PB. Population two takes over and the dynamics relax back, maintaining item two in WM via the persistent activity 
(as shown in column (A) of \red{Fig.} \ref{Sfig1}).  
For $\Delta T_2 \ge 130$ ms we observe the emission of two or more successive PBs in population two during the stimulation period. These PBs noticeably depress $x_2$, which is unable to recover before population one, characterized by a larger amount of available resources $x_1$, emits a PB and item one takes over (see column (B) of \red{Fig.} \ref{Sfig1}). However, for much longer $\Delta T_2 \ge 850$ ms, the absence of activity in population one leads to a noticeable decrease of the utilization factor $u_1$.  As a matter of fact at the end of the stimulation period, the item two is finally selected (as shown in column (C) of \red{Fig.} \ref{Sfig1}). 

To summarize, item two can be selected when the duration of the stimulation is restricted to a narrow time interval,
sufficiently long to render the efferent synapses of population two stronger than those of population one,
but not so long to highly deplete the resources of the same synapses. Furthermore, item two can be selected 
when $\Delta T_2$ is sufficiently long to allow for the decay of the synaptic facilitation of population one
towards its baseline value.

%\clearpage
\subsection*{Multi-Item Memory Loading}
\label{7items}

\begin{figure}[!h]
	\begin{adjustwidth}{-1.25in}{0in}
	\includegraphics[width=1\linewidth]{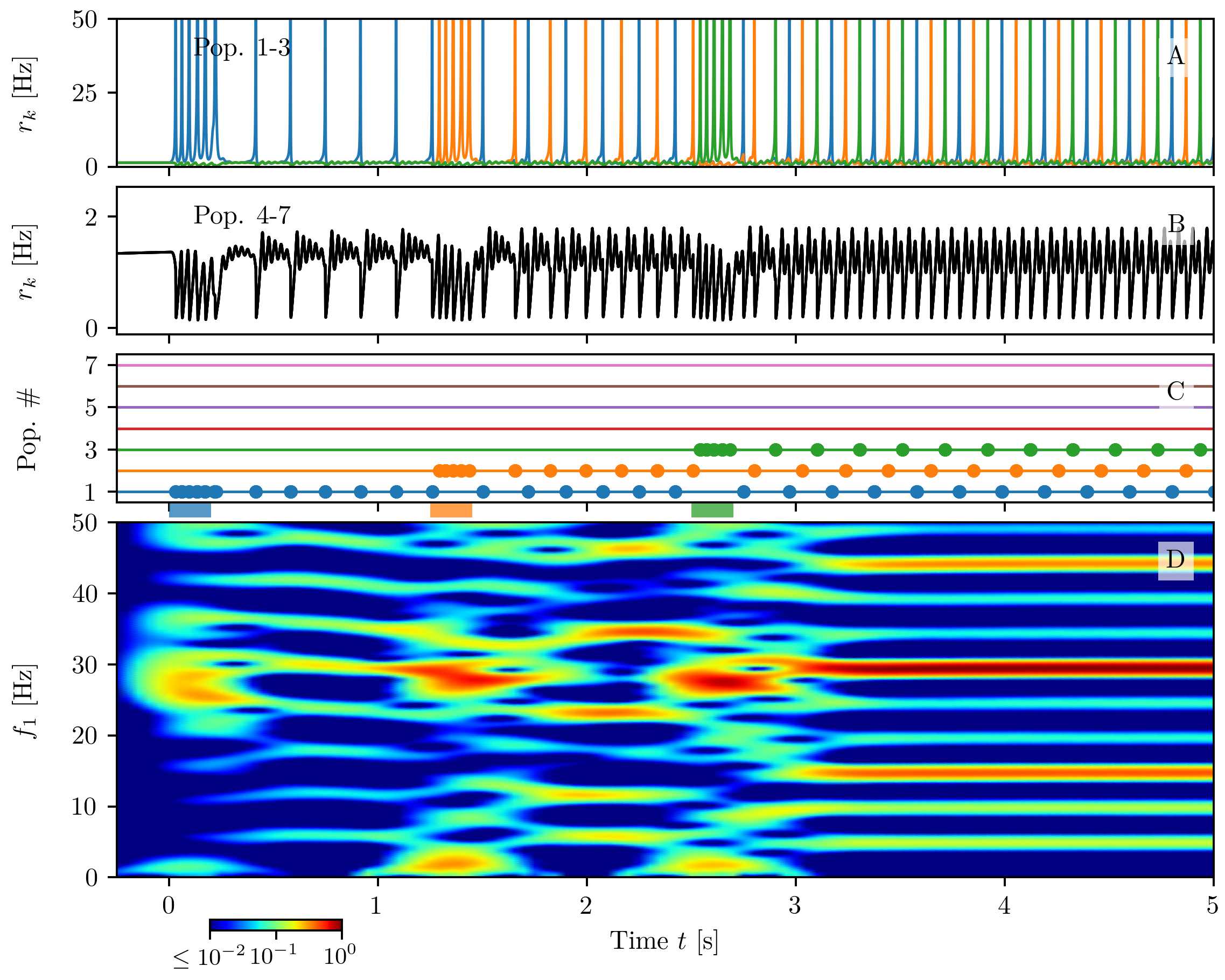}
     \caption{{\bf Multi-item memory loading} Firing rates $r_k(t)$ of excitatory populations: blue, orange and green curves corresponds to $k=1,2,3$ ({\bf A}), while the black curve refers to $k=4,\dots,7$ ({\bf B}). PBs of excitatory populations are
     shown in ({\bf C}): horizontal lines in absence of dots indicate low activity regimes; coloured dots mark the PBs' occurrences.
      The coloured bars on the time axis denote the presence of a stimulation pulse targeting the corresponding population. Only populations $k=1,\dots, 3$ are stimulated in the present example. The spectrogram of the average membrane potential
      of population one $v_1(t)$ is reported in panel ({\bf D}). 
Parameters: $\Npop = 7$,  $\taum^\mathrm{e}=15$ ms, $\taum^\mathrm{i}=10$ ms, $\Jeep=154$, $\Jeeb=\frac{4}{7}\cdot18.5$, $\Jei=-26$, $\Jie=\frac{4}{7}\cdot 97$, $\Jii=-60$, $\Iback=0$, $H^\mathrm{(e)}=0.05$, $H^\mathrm{(i)}=-2$, $\Delta^{(\mathrm{e})}=\Delta^{(\mathrm{i})}=0.1$} 
	\label{fig:7item_setup_load3}
	\end{adjustwidth}
\end{figure}

In order to be able to load a higher number of memory items in WM, we consider the neural mass model
(Eqs. \eqref{eq:fre}) with m-STP (Eqs. \eqref{eq:dumeanfied}) arranged in a  more complex
architecture composed by $\Npop=7$ excitatory and one inhibitory populations,
each excitatory population coding for one memory item. 
The system is initialized with all the populations in the silent state. Then, each item is loaded 
by delivering an excitatory pulse of amplitude $\Delta I^{(k)}=1$ and duration of 0.2 s to the chosen population, while successive items 
are loaded at intervals of 1.25 s, as shown in Fig. \ref{fig:7item_setup_load3} (C).

Let us first consider the successive loading of $N_L=3$ items shown in Fig. \ref{fig:7item_setup_load3}.
As one can appreciate from the spectrogram reported in panel (D), during the loading phase,
each stimulated excitatory population emits a  sequence of PBs in the $\beta-\gamma$ range
($\simeq 27$ Hz) via a PING-like mechanism mediated by the inhibitory population. This is
joined to stimulus locked transient oscillations in the $\delta$-band around 2 Hz.
Furthermore, during each loading period, the activity of the other populations is interrupted
and it recovers when the stimulation ends. 
These results resemble the LFP measurement performed in PFC of primates 
during coding of objects in short term-memory \cite{siegel2009phase,lundqvist2016gamma}. In particular,
the experimentally measured power spectrum of the LFP displays transient oscillations at $\simeq 2-4$ Hz in the 
$\delta$ range, phase-locked to stimulus presentation, together with tonic oscillations at $\simeq 32$ Hz.

As shown in Fig. \ref{fig:7item_setup_load3} (A), the loading of each item is followed by the emission of PBs 
from the stimulated population at regular time intervals $T_\mathrm{c}$.
\red{Therefore, in this case, the number $N_I$ of items retained in the WM coincides with the number $N_L$ of the loaded ones}. The PBs of all the excited populations arrange in a sort of {\it splay state}  with inter-burst
periods $T_\mathrm{b} = T_\mathrm{c}/N_\mathrm{I}$ \cite{olmi2012}. The period $T_\mathrm{c}$ depends on the
number of retained items: for $N_\mathrm{I}=3$ we have $T_\mathrm{c} \simeq 0.2035$ s.

The non stimulated populations display a low firing activity modulated by the slow PB emission occurring 
in the excitatory populations and by the fast $\beta-\gamma$ \red{transient oscillations towards the focus equilibrium, sustained
by the inhibitory activity (see Fig. \ref{fig:7item_setup_load3} (B)). In particular, one observes fast 
COs ($\simeq 27$ Hz) nested in slow oscillations, characterized by a frequency increasing with the
number of loaded items (from the $\theta$-band for 1 item to the $\beta$-band for 3 items), 
somehow similarly to what shown in \cite{lisman1995,jensen1996novel,kopell2011}.}
 
A more detailed analysis of the spectrogram in Fig. \ref{fig:7item_setup_load3} (D) reveals that, after each loading phase, several harmonics of the fundamental frequency $f_c \equiv 1/T_\mathrm{c} \simeq 5$ Hz are excited. In particular after the complete loading
of the three items the most enhanced harmonics are those corresponding to $3 f_c \simeq 15 $ Hz  and $6 f_c \simeq 30$ Hz.
This is due to the coincidence of the frequency $3 f_c = f_\mathrm{b} \equiv 1/T_\mathrm{b}$ with the inter-burst frequency $f_\mathrm{b}$, while the main peak in the spectrogram,
located at $30$ Hz, is particularly enhanced due to the resonance with the proximal $\beta-\gamma$ 
rhythm associated to the \red{PING-like oscillations emerging during memory loading.}

\begin{figure}[!h]
	\begin{adjustwidth}{-1.25in}{0in}
	\includegraphics[width=1\linewidth]{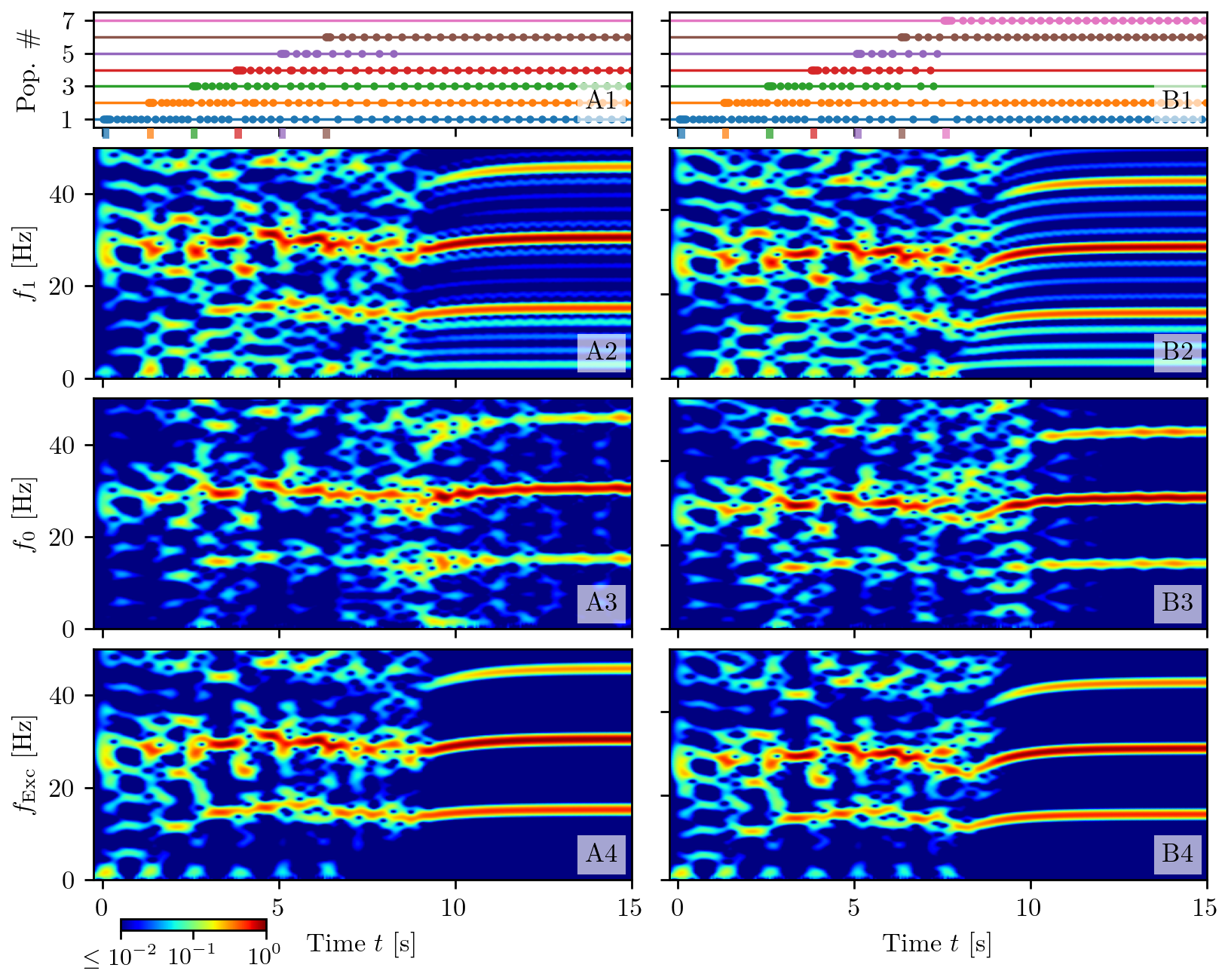}
	\caption{{\bf Maximal capacity.} \red{Response of the system when $N_L=6$ (column (A)) or $N_L=7$ (column (B)) excitatory populations are successively stimulated at a presentation rate of 0.8 Hz}. 	
	Population bursts of excitatory populations \red{({\bf A1-B1})}: horizontal lines in absence of dots indicate quiescence phases at low firing rates $r_k$ for populations $k=1,...7$. Dots mark PBs of the corresponding population. The coloured bars on the time axis mark the starting and ending time of stimulating pulses, targeting each population. 
Spectrograms of the mean membrane potential $v_1(t)$ \red{({\bf A2-B2})}, $v_0(t)$ \red{({\bf A3-B3})} and of the mean membrane potentials averaged over all the excitatory populations \red{({\bf A4-B4})}; \red{for clarity the corresponding frequencies have been denoted as $f_1$, $f_0$ and $f_\mathrm{Exc}$}. Parameter values as in Fig. \ref{fig:7item_setup_load3}.}  
	\label{fig:7item_setup_load6}
	\end{adjustwidth}
\end{figure}

It is interesting to notice that the transient $\delta$ oscillations related to the stimulation phase are only supported by the excitatory population dynamics, as evident by comparing the spectrograms for $v_1(t)$ and for the membrane potential averaged over all excitatory populations (shown in panels \red{(A2-B2)} and \red{(A4-B4)} of Fig. \ref{fig:7item_setup_load6}, \red{(B) and (D) of \nameref{Sfig2}}) with those for the inhibitory population (shown in panel (A3-B3) of Fig. \ref{fig:7item_setup_load6}, \red{and (C) of \nameref{Sfig2}}). Another important aspect is that only the harmonics of $f_\mathrm{b} \equiv N_\mathrm{I}f_c $
are present in the spectrogram of the inhibitory population, and in that associated to the average excitatory membrane potential, 
but not all the other harmonics of $f_c$. This is related to the fact that, in the average of 
the mean excitatory membrane potentials, the bursts of all populations will be present with a period $T_\mathrm{b}$, therefore in this
time signal there is no more trace of the periodic activity of the single population. This is due to the fact that
$T_\mathrm{c} \equiv N_\mathrm{I} T_\mathrm{b}$.  Furthermore, in the dynamics of the inhibitory population, the activity of all excitatory populations are reflected with the same weight, therefore also in this case there is no sign of the fundamental frequency $f_c$.

\red{
\subsubsection*{Memory Capacity}}
 
If we continue to load more items, we discover that up to five items are loaded
and retained by WM (see \nameref{Sfig2} (A)). However, if we load a sixth
item, WM is able to maintain all the loaded $N_\mathrm{L}=6$ items only for a short time interval $\simeq 1$ s, after which 
population five stops delivering further PBs. The corresponding item is not in WM anymore and the memory load returns to $N_\mathrm{I}=5$ (as shown in Fig. \ref{fig:7item_setup_load6} (A)).
By loading $N_L=7$ items we can induce more complex instabilities in the WM, indeed the experiment reported in
Fig. \ref{fig:7item_setup_load6} \red{column (B)} reveals that, in the end, only $N_\mathrm{I}=4$ items can be maintained, suggesting that a too fast acquisition of new memory items can compromise also already stored WM items.
As a matter of fact, only the oldest and newest items are retained by WM:
this in an example of the so-called {\it primacy and recency effect}, which 
has been reported in many contexts when a list of items should be memorized \cite{postman1965,morrison2014}.

\begin{figure}[!h]
	\includegraphics[width=1\linewidth]{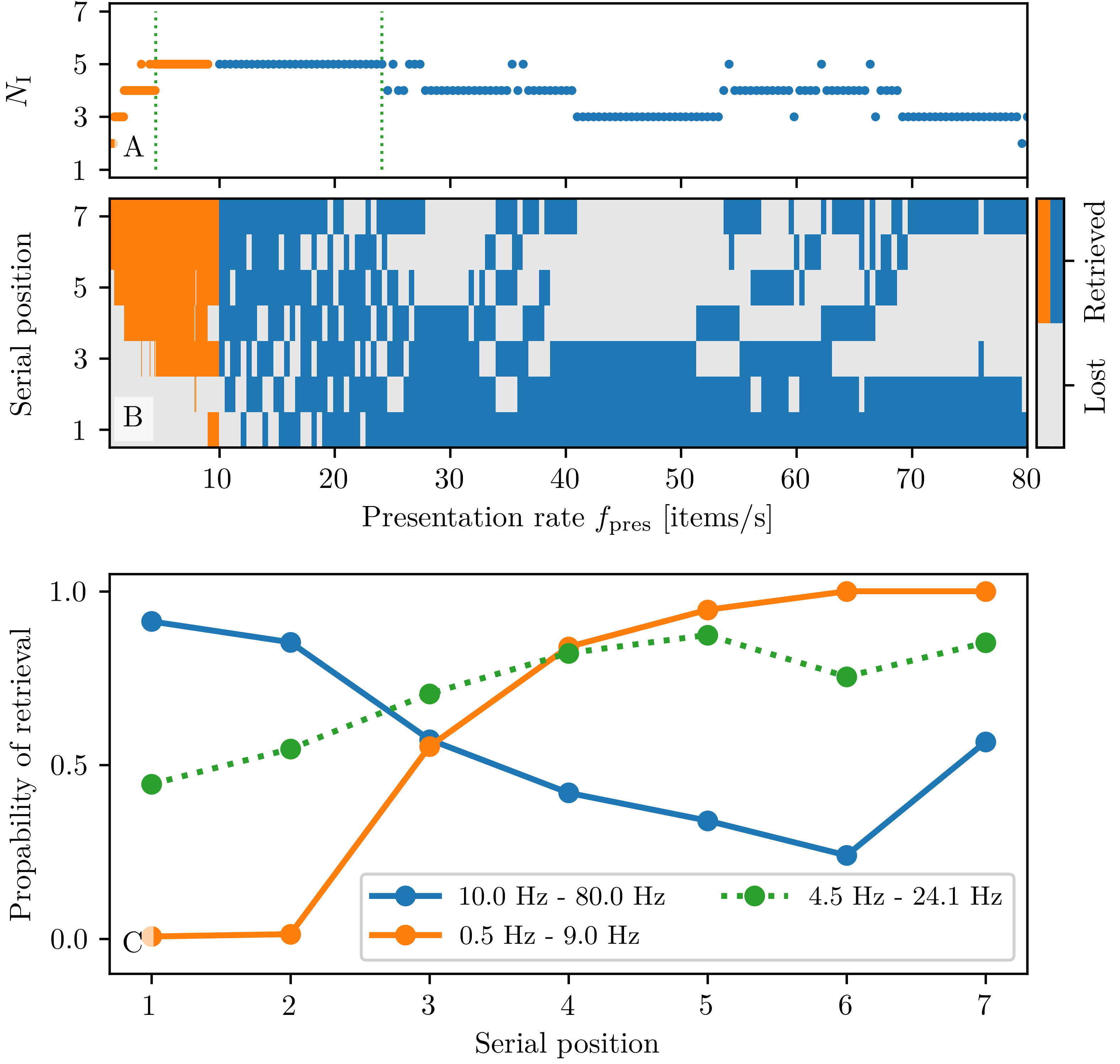}
	\caption{\red{{\bf Dependence of the memory capacity on the presentation rate.}  
    {\bf (A)} Total number of retrieved items $N_\mathrm{I}$ vs. presentation frequency $f_\mathrm{pres}$ for slow ($[0.5 : 9.0]$ Hz, orange) and fast presentation rates ($[10 : 80]$ Hz, blue). The two green dotted vertical lines denote an optimal rate interval where $N_\mathrm{I}$ is maximal.  {\bf (B)} Map showing the retrieved items at a given serial position for presentation rates $f_\mathrm{pres}$ in the interval $[0.5:80]$ Hz.  {\bf (C)} Probability of retrieval vs. serial position for slow (orange) and fast (blue) presentation rates. In order to estimate these probabilities 150 equidistant rates are considered in the interval $[0.5 : 9]$ Hz and 150 in $[10, 80]$ Hz. The green dotted line refers to the probability estimated within the rate interval enclosed between two green dotted vertical lines in (A).
An item is considered as retrieved if the corresponding population is still delivering PBs 20 s after 
the last stimulation. All other parameter values are as in Fig. \ref{fig:7item_setup_load3}.}}
	\label{fig:7item_pres_freq_scan}
\end{figure}

\red{To investigate more in detail the memory capacity of our model, we have considered different presentation rates $f_{\rm{pres}}$ for the items. In particular, we
have sequentially delivered a stimulation pulse to all the excitatory populations,
from the first to the seventh, with a presentation rate $f_\mathrm{pres}$; each pulse is characterized by an amplitude  $\Delta I=16$ and a time width $\Delta T = 1/f_\mathrm{pres}$. Results are presented in Fig. \ref{fig:7item_pres_freq_scan} for presentation rates in the interval $[0.5 : 80]$ Hz. From panel (A) it turns out that the maximal capacity is five, a value that can be mostly achieved for presentation rates within an optimal range $f_{\rm pres} \in [4.5:24.1]$ Hz, delimited for better clarity by green dashed lines in panel (A). For these {\it optimal} rates we have
estimated the probability of retrieving a certain item versus its presentation position, as shown in
panel (C) (green dashed line). The probability displays very limited variations ($\simeq 45-85$ \%) for
the 7 considered items, suggesting that in this range the retrivial of an item does not strongly depend on its position in the presentation sequence. Conversely, for slow rates ($f_{\rm{pres}} \leq 9$ Hz) the last loaded items result to be the retained ones
({\it recency effect}), as evident from the orange symbols in panel (B). This is confirmed by the calculation of
the probability of retrieval versus the corresponding serial position shown in panel (C) (orange curve), obtained by considering 150 equidistant rate values in the interval $[0.5:9]$ Hz. Finally, the {\it primacy and recency effect} is observable only for presentation rates faster than $9$ Hz (see blue symbols in panel (B)), as confirmed by the probability 
shown in panel (C) (blue curve), which has been obtained by considering 150 equidistant frequencies within the interval $[10: 80]$ Hz. 
}

A theoretical estimation of the maximal capacity $N_\mathrm{c}^{\mathrm{max}}$ for WM, based on short-term synaptic plasticity, has been recently derived in \cite{mi2017synaptic}. \red{By following the approach outlined in  \cite{mi2017synaptic}, we have been able to obtain an analytical expression for $N_\mathrm{c}^{\mathrm{max}}$
also for our neural mass model, namely }
\begin{equation}
N_\mathrm{c}^{\mathrm{max}} \simeq  \frac{\taud}{\taum^\mathrm{e}} \ln{\left[\frac{\tauf/\taud}{1-U_0}\right]} \frac{\sqrt{C}}{\pi}
\quad ;
\label{maxmem1}
\end{equation}
where \red{$C=\left[H^{(\mathrm{e})} +\Iback + \taum^{\rm e} \left(-|J_{\rm ei}|+ {\bar J} \right) \frac{\sqrt{H^{(\mathrm{e})} + \Iback}}{\pi} \right]$, with ${\bar J} = [\Jeep + (\Npop-2)\Jeeb] {\bar x} {\bar u} $. The explicit derivation of \eqref{maxmem1} is reported} in \nameref{sec:capacity}. 
As shown in \cite{mi2017synaptic}, the value of $N_\mathrm{c}^{\mathrm{max}}$ is essentially dictated by
the recovery time of the synaptic resources $\taud$ and has a weaker (logarithmic) dependence on the 
ratio between the facilitation and depression time scales. Furthermore, for our model, the maximal capacity increases
by increasing $H^\mathrm{(e)}$, $\Iback$ and the self and cross excitatory synaptic couplings, while it decreases 
whenever the coupling from inhibitory to excitatory population is \red{strengthened}.

By employing the theoretical estimation \eqref{maxmem1} we get $N_\mathrm{c}^{\mathrm{max}}  \in [3.6, 4.8]$, in pretty good agreement
with the measured maximal capacity. We can affirm this in view of the results reported in \cite{mi2017synaptic} for a different mean field model, where  the analytical predictions  overestimate the maximal capacity by a factor two.

\red{
\subsubsection*{Memory Load Characterization}
}

A feature usually investigated in experiments during the loading or juggling of more items is
the power associated to different frequency bands. In particular, a recent experiment has examined
the frequency spectra of LFPs measured from cortical areas of monkeys while they maintained
multiple visual stimuli in WM \cite{kornblith2016}. These results have revealed that 
higher-frequency power (50–100 Hz) increased with the number of loaded stimuli, while
lower-frequency power (8–50 Hz) decreased. 
%More detaled characterizations of the 
Furthermore, the analysis of a detailed network model, biophysically inspired by the PFC structure, has
shown that $\theta$ and $\gamma$ power increased with memory load, while
the power in the $\alpha$-$\beta$ band decreased \cite{lundqvist2011theta},
in accordance with experimental findings in monkeys \cite{wimmer2016}. \red{However,
for humans, an enhance in the oscillatory power during WM retention has been reported for
$\theta$ \cite{gevins1997,jensen2002}, $\beta$ and $\gamma$-bands \cite{tallon1998,howard2003gamma},
while no relevant variation has been registered in the $\alpha$-band \cite{tallon1998}.}

\begin{figure}[!h]
	\includegraphics[width=1\linewidth]{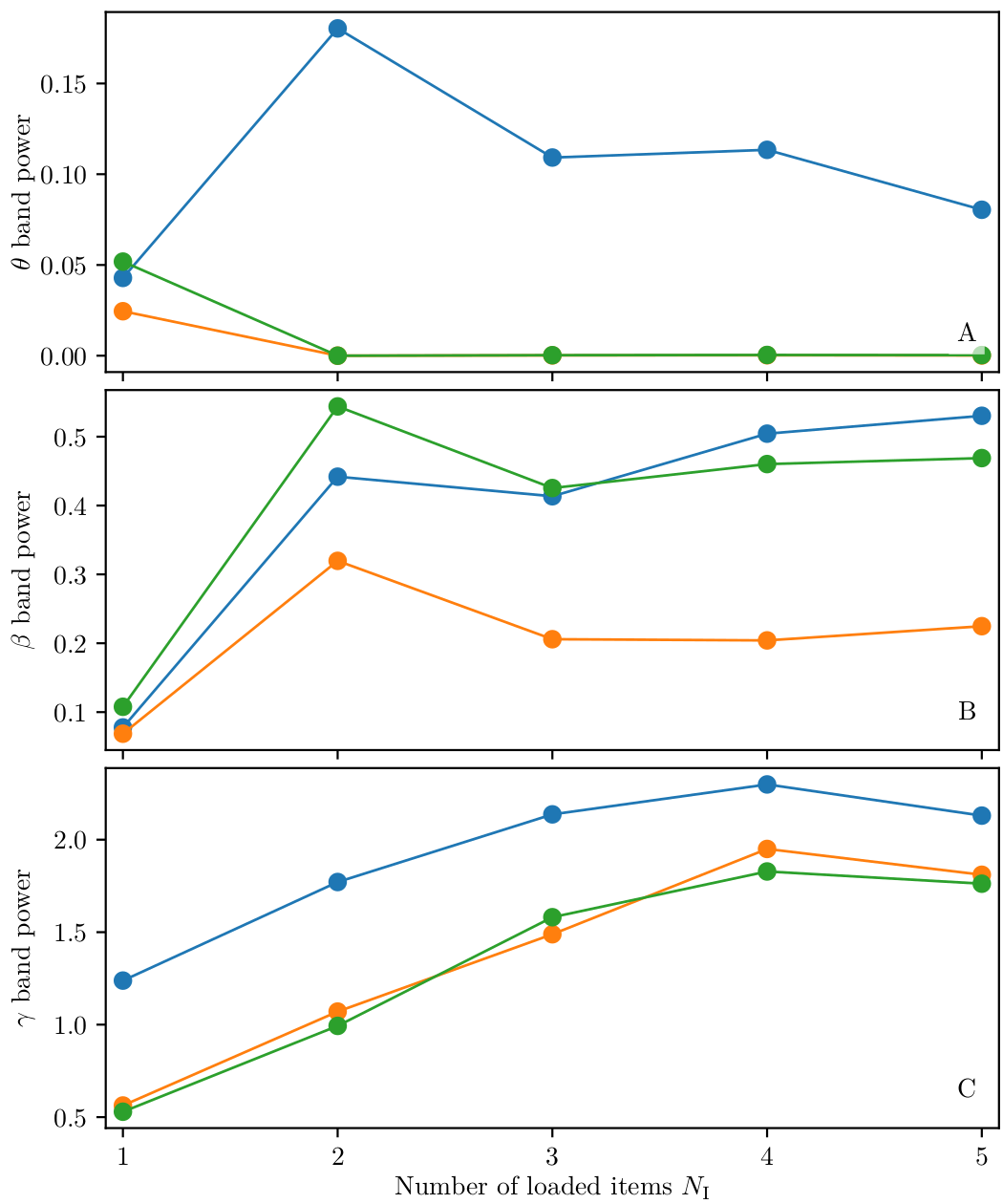}
	\caption{{\bf Dependence of the power on the number of loaded items} 
Power in the \red{$\theta$-band (3-11 Hz) ({\bf A}) , in the $\beta$-band (11-25 Hz) ({\bf B})} and in the $\gamma$-band (25-100 Hz) ({\bf C}) as a function of the number $N_\mathrm{L}=N_I \le 5$ of loaded items. The integral of the power spectra in the specified frequency bands are displayed for the mean membrane potential $v_1(t)$ of the excitatory population one (blue symbols), the mean membrane potential $v_0(t)$ of the inhibitory population (orange symbols) and the mean membrane potentials averaged over all the excitatory populations (green symbols). The power spectra have been evaluated
over a 10 s time window, after the loading of $N_\mathrm{I}$ items, when these items were juggling in WM. 
Parameter values as in Fig. \ref{fig:7item_setup_load3}.}
	\label{fig:power}
\end{figure}

We considered the loading of $N_\mathrm{L} \le 5$ items and estimated the corresponding power spectra,
after the loading of all the considered items, for the following variables: 
i) the mean membrane potential $v_1(t)$ of the excitatory population one;
ii) the mean membrane potential $v_0$ of the inhibitory population; iii) the mean membrane potential averaged over
all the excitatory populations. The power has been estimated in the $\theta$, $\beta$ and
$\gamma$-bands by integrating the spectra in such frequency intervals.
The results for $ 1 \le N_\mathrm{I} \equiv N_L \le 5$ shown in Fig. \ref{fig:power} reveal that the power in the $\gamma$-band 
is essentially increasing with $N_\mathrm{I}$ for all the considered signals, as reported experimentally for several brain areas involved in WM \cite{howard2003gamma,van2010hippocampal,roux2012gamma}. 
\red{Furthermore, the $\gamma$-power obtained for $v_0$ and the mean excitatory potential are
essentially coincident (see panel (C)), thus confirming the fundamental role
of the inhibition in sustaining $\gamma$ oscillations via a PING mechanism.
Moreover, the power in the $\beta$-band, reported in panel (B), increases almost monotonically for $v_1$, while it displays a non monotonic behaviour for the inhibitory population and for the average excitatory membrane potential, while saturating to a constant value for $N_I \ge 3$. More striking differences emerge when considering the $\theta$ power (shown in panel (A)): in this case the power spectra for $v_0$ and the average excitatory membrane potential display almost no variations with $N_I$,
while that for $v_1$ increases by passing from one to two loaded items before saturating for larger $N_I$.
The observed discrepancies can be explained by the fact that the fundamental frequency $f_c$ and its
harmonics are present in the spectrum of $v_1$, while they are absent both in the spectrum of $v_0$ and of the average excitatory membrane potential.
We have not observed any variation in the $\alpha$-power analogously to what reported experimentally in \cite{tallon1998}.
}

\red{Finally, inspired by a series of experimental works reporting neurophysiological measures of
WM capacity in humans \cite{vogel2004neural,vogel2005neural}, we have investigated
if a similar indicator can be defined also in our context. In particular, the authors
in  \cite{vogel2004neural} measured, as a neural correlate of visual memory capacity,
the event-related potentials (ERPs) from normal young adults performing a visual memory task.
To each patient was presented a bilateral array of $2 \times N_L$ coloured squares and he/she was
asked to remember the $N_L$ items in only one of the two hemifields. Due to the organization of the visual system,
the relevant ERPs associated to this visual stimulation should appear in the controlateral hemisphere.
Therefore the difference  between controlateral and ipsilateral activity has been measured in order to remove
any nonspecific bilateral ERP activity. The authors observed that, by increasing the number of squares $N_L$, also the
ERP difference increases, while it saturates by approaching the maximal capacity (measured during the same test)
and even decreases for $N_L > N_c^{max}$. Thus the ERP difference can be employed as a reliable neurophysiological predictor
of memory capacity.}

\begin{figure}[!h]
	\includegraphics[width=1\linewidth]{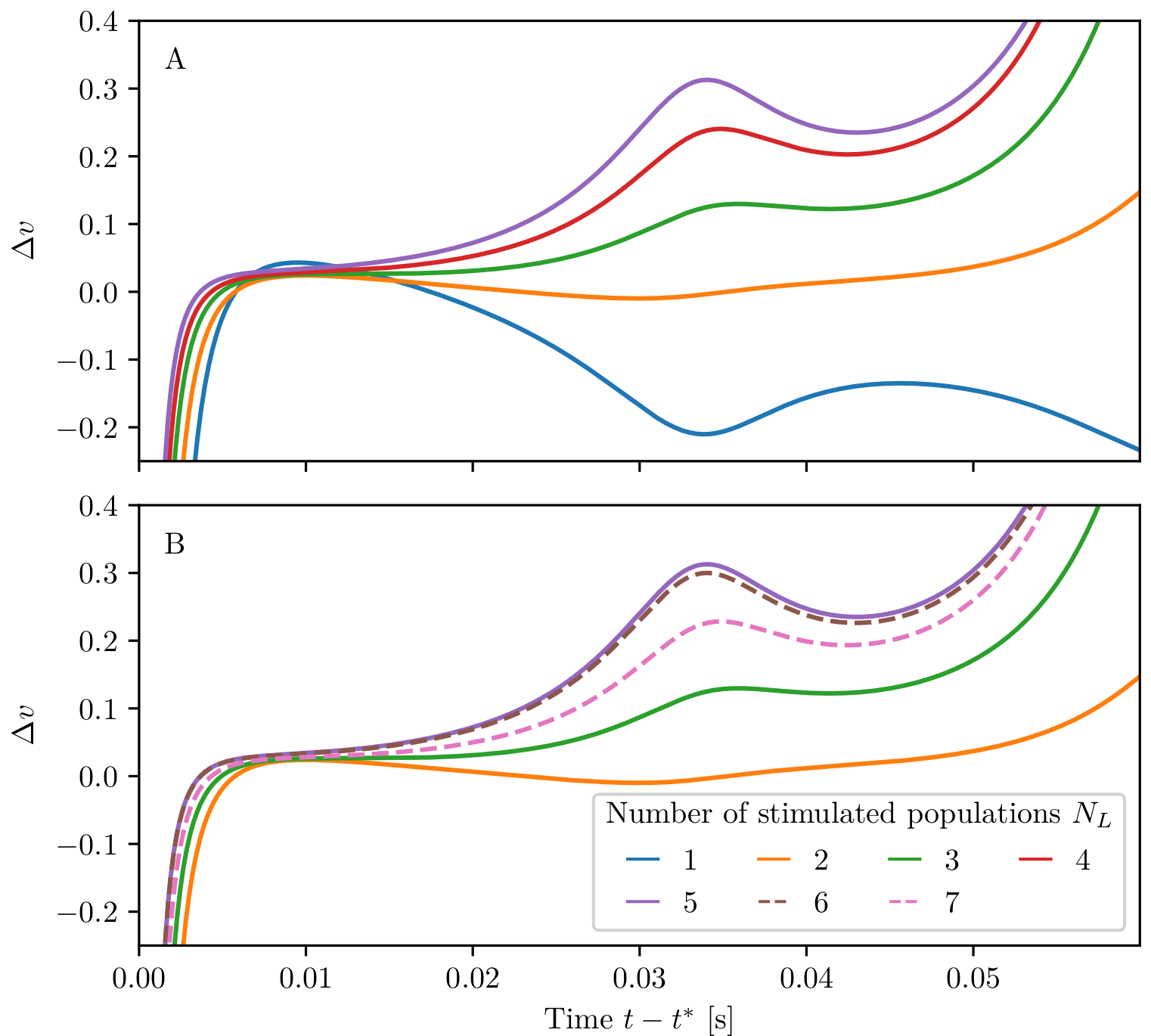}
	\caption{\red{{\bf Dependence of the membrane potential difference on the number of loaded items.} 
	Difference $\Delta v$ of membrane potentials 
	versus time when stimulating different numbers $N_L$ of populations. The presented time-series are aligned to $t^\ast$, which marks the deliverance of a PB from population one.
The populations are stimulated sequentially with parameter values as in Fig. \ref{fig:7item_setup_load3}.}}
	\label{fig:saturation}
\end{figure}

\red{In order to define a similar indicator in our case, we have calculated the membrane potential difference $\Delta v$
between the mean membrane potential, averaged over the populations coding, for the retained items 
(whose activity is reported for example in Fig. \ref{fig:7item_setup_load3} (A)) and the one averaged
over the non-coding populations (e.g. see Fig. \ref{fig:7item_setup_load3} (B)).
By measuring $\Delta v$ just after the deliverance of a PB by the first population,
we observe a clear growth of this quantity with the number $N_L$ of presented items,
as shown in Fig. \ref{fig:saturation} (A) for $N_L=1,2,3,4,5$. However, as soon as $N_L > N_c^{max}$ the
membrane difference $\Delta v$ almost saturates to the profile attained for $N_L = N_c^{max} =5$
and even decreases for larger $N_L$, as evident from Fig. \ref{fig:saturation} (B) where
we report the results also for $N_L=6$ and  7 (dashed lines). Analogous results have been obtained
by considering the mean membrane potential of the next to fire populations instead of 
the difference $\Delta v$. Therefore, the results are not biased by the chosen indicator
and $\Delta v$ allows for a better presentation clarity.
}

\red{To conclude this sub-section, we can affirm that the mean membrane potential can be employed,
analogously to the ERP in the experiments \cite{vogel2004neural,vogel2005neural}, as a proxy
to measure the memory load and capacity. It is worth noticing that neither the mean membrane potential nor the ERP
are accessible for firing rate models.}

\section*{Discussion}
 
In this paper we have \red{introduced} a next generation neural mass model
for WM based on short-term depression and facilitation.
\red{The model has been developed to reproduce exactly the macroscopic dynamics 
of heterogeneous QIF spiking neural networks with mesoscopic short-term plasticity
(m-STP), in the limit of an infinite number of neurons with Lorentzian distributed 
excitabilities. Even though the choice of the excitability distribution allows for an analytical derivation of the 
model, it does not limit the generality of the results \cite{montbrio2015macroscopic, bi2020}.
As shown in sub-section \nameref{sec:comparison}, the neural mass model reproduces well not only the QIF network dynamics
with m-STP plasticity but, to a large extent, it reproduces also the dynamics of networks with plasticity 
implemented at a microscopic level ($\upmu$-STP). Therefore, the macroscopic dynamics of 
spiking neural networks made of hundreds of thousands neurons with 
synapses evolving accordingly to realistic cellular mechanisms,
can be well captured in terms of a four dimensional mean-field model.
}

The novelty of this neural mass model, besides not being heuristic,
but derived in an exact manner from the microscopic underlying dynamics,
is that it reproduces the evolution of the population
firing rate as well as of the mean membrane potential. This allows us
to get insight not only on the synchronized spiking activity, but also on 
the sub-threshold dynamics and to extract information correlated to LFPs,
\red{EEGs and ERPs, that are usually measured during WM tasks to
characterize the activity of the brain at a mesoscopic/macroscopic scale.} 
\red{The knowledge of the mean membrane potential evolution is fundamental in order to capture the
dynamics displayed by the underlying microscopic QIF network, already in the absence
of plasticity. Indeed, at variance with this next generation of neural mass models, a rate model cannot 
reproduce neither fast oscillations observed in purely inhibitory networks \cite{devalle2017firing}, nor memory clearance obtained via a resonance mechanism
between an external $\beta$ forcing and the intrinsic network oscillations\cite{schmidt2018network}.}
\red{As we have shown in the sub-section \nameref{comparison_fr}, a heuristic firing rate model,
specifically designed to reproduce the QIF network dynamics with STP, 
does not display any oscillatory activity in the $\beta$-$\gamma$ range,
contrary to what observable in the spiking network itself and in our neural mass model. It should be
remarked that this represents a main drawback when employing firing rate models to mimic WM operations,
because the emergence of $\beta$-$\gamma$ rhythms is intimately related to short-term memory activity, 
as shown in several experiments on humans and monkeys \cite{tallon1998,pesaran2002,howard2003gamma,siegel2009phase, lundqvist2016gamma, wimmer2016}.}

The macroscopic model we developed is extremely flexible since, depending on the operational point, 
it can mimic WM  maintenance in terms of persistent spiking activity or in terms
of \red{selective and spontaneous reactivation, based on 
plastic footprints into the synapses. In particular, short-term depression and facilitation complement 
each other to obtain an efficient WM model based on synaptic reactivation. Facilitation allows us to maintain 
trace of the loaded items stored in WM, while depression is responsible for the bursting activity that 
refreshes the WM content.}

Memory loading is characterized, \red{in any operational condition}, 
by collective oscillations (PBs) with frequencies $\simeq 22 - 28$ Hz  in the
$\beta$-$\gamma$ band. These oscillations emerge \red{spontaneously}
in the considered model thanks to a \red{PING-like mechanism
triggered by the transient oscillations of the selectively stimulated excitatory population
towards a focus equilibrium and sustained by the common inhibitory pool. The memory loading 
induces also stimulus-locked transient oscillations involving
the harmonics of a fundamental frequency (i.e. $\simeq 2-3$ Hz).
These results strongly resemble the spectral features observed in 
experiments on humans and monkeys performing WM tasks. Of particular interest
are the responses of the primary somatosensory cortex in humans
to vibrotactile stimuli \cite{spitzer2010} and the population activity of the PFC measured during 
a behavioural task of object recognition, performed by monkeys \cite{siegel2009phase,lundqvist2016gamma}.
In the first experiment stimulus-locked EEG signals have been measured
revealing a transient broad-band activity in the 4-15 Hz
range, followed by a stationary activity at $\simeq 26$ Hz 
lasting for the duration of the stimulus \cite{spitzer2010}. These
findings are quite similar to the spectral features displayed by our model 
during one memory item loading, see Fig.  \ref{fig0}.
In the experiments on primates, the analysis of the LFP spectrograms revealed an evoked response around 2-4 Hz and tonic
oscillations around 32 Hz \cite{siegel2009phase,lundqvist2016gamma}, thus resembling
the power spectra that we have obtained for multi-item memory loading reported in Figs. 
\ref{fig:7item_setup_load3} and \ref{fig:7item_setup_load6}.
}

Memory maintenance in the synaptic-based model is ensured
by facilitation over a time scale of a few seconds, even in absence of spiking activity. The memory
can be refreshed, and therefore maintained for a longer period,
thanks to the reactivation of the synaptic resources of the excitatory population storing the item.
This can be obtained, on one side, via brief non specific stimulations  given to
all the excitatory populations, analogously to the reactivation of latent working memories performed 
with single-pulse transcranial magnetic stimulation realised in humans \cite{rose2016}.
In particular, to maintain the memory for long time intervals, the stimulations
should be delivered with a period smaller than the decay time of the facilitation.
On the other side the memory maintenance over long periods can be
achieved thanks to the spontaneous emergence of periodic PBs delivered by the
excitatory population coding for the loaded item.

An interesting aspect that we have investigated, concerns
the competition between two items loaded in non-overlapping populations.
When the WM load consists already of one item coded by an excitatory population, 
the stimulation of the other population can be regarded as a distractor.
If the items are stored as repetitive PBs, we can observe three outcomes: 
for brief stimulations the distractor has no influence on WM, and the first item
remains loaded; for sufficiently long stimulations the second item is loaded in WM;
for intermediate situations both items are maintained in WM.
In this latter case both populations deliver periodic trains of PBs 
arranged in anti-phase. Such behaviour can be seen as a neural correlate to two 
items juggling in WM \cite{lewis2014}.
For sufficiently long and strong stimulations  
one observe a chaotic scenario \cite{cortes2013}, where the final outcome
can be any of the three described above and it
depends on small differences in the perturbation deliverance. 
Our findings can help in elucidating the results reported in
\cite{lewis2014}, where it has been shown that, when an attended and an
unattended item are juggling for a long period, the unattended item
can prevail leading to a loss of the stored memory.

If the items are stored in WM as persistent states, the memory juggling
is no more observable. As in the previous case, for sufficiently long 
stimulations, the second item substitutes the first one in WM. 
However, due to synaptic depression and facilitation, even a short stimulation
can lead to the loading of the second item in WM, whenever its duration
falls in a narrow time interval ($\simeq {\cal O}(\taud/2)$) and the amplitude of 
the stimulation is sufficiently large. This suggests that there are optimal
stimulation strategies to ensure a fast learning of a new item in WM.

\red{By considering a neural architecture composed of multiple excitatory populations 
and a common inhibitory pool, more items can be maintained at the same
time in WM as periodic trains of PBs. This is observable despite the possible interference among excitatory
populations that are allowed to directly interact, thanks to the couplings present among them,
and not only via the inhibitory pool, as in similar architectures
considered in the literature \cite{mi2017synaptic}.}  
All populations coding for stored items follow the same periodic dynamics, but they
deliver PBs at evenly shifted phases, similarly to the
splay states observed for globally coupled excitatory neuronal networks \cite{olmi2012}.
The inter-burst interval of two successive PBs approaches a value
\mbox{$T_\mathrm{b} \simeq 65$ ms} when more than 2 items are loaded. This clearly
induces the emergence of a peak in the power spectra of the mean membrane
potentials in the $\beta$ band for $f_\mathrm{b} \simeq 15$ Hz. However, the most prominent
peak in the spectra is around 30 Hz, due to the resonance of the second
harmonic of $f_\mathrm{b}$ with the oscillations of the inhibitory population
in the $\beta-\gamma$ range. These oscillations are associated \red{to damped oscillations
of the excitatory population towards a focus equilibrium, induced by the loading of one memory item  
and sustained by the inhibitory pool thanks to a PING-like mechanism}.
\red{The considered architecture allows for the maintenance of multi-items in WM
thanks to their spontaneous reactivation}, without destructive or interference effects often reported for models
based on persistent spiking \cite{miller2018working}. Furthermore, the memory of 
different items is associated to preferential phases with respect to the 
collective limit cycle behaviour of the whole system, characterized by a period $T_\mathrm{c} = N_\mathrm{I} T_\mathrm{b}$,
where $N_\mathrm{I}$ is the number of retained items. Experimental evidences of phase-dependent 
neural coding of objects in the PFC of monkeys have been reported in \cite{siegel2009phase}.

\red{The memory load $N_I$ of our model depends on the presentation rate $f_{\rm pres}$
of a sequence of memory items, however it is always limited between $3 \le  N_I \le 5$
analogously to what reported in many analysis concerning the WM capacity \cite{cowan2001,cowan2010magical}.
In particular, for slow presentation rates ($f_{\rm pres} \le 8$ Hz) we observe that $N_I$ grows
proportionally to $f_{\rm pres}$ and that only the last presented items are retained in the memory.
The maximal capacity $N_c^{max} = 5$ can be attained mainly within an optimal range of presentation rates,
namely $[4.5:21.4]$ Hz. These rates correspond to the characteristic frequencies associated
to the PB dynamics of the model, since the inter-burst frequency is $f_\mathrm{b} \simeq 12-16$ Hz for $N_I > 2$,
while the oscillation frequency of a single population is $f_\mathrm{c} \simeq 3 - 5$ Hz. In this
optimal range there is no clear preference for the item retained in the memory and its
serial position in the loaded sequence. For faster frequencies $f_{\rm pres} > 25$ Hz, a destructive
interference among the items leads to a decrease in $N_I$: this mechanism has been suggested in 
\cite{cowan2010magical} to be at the origin of the reduced capacity. For sufficiently fast
rates $f_{\rm pres} \ge 10$ Hz, a {\it primacy and recency effect} \cite{postman1965,morrison2014}
is observable with a prevalence for the first loaded items to be retained.}

\red{To obtain a better understanding of the capacity limits of our model,
we have derived an analytical expression for the maximal capacity $N_c^{max}$ by following
the approach outlined in \cite{mi2017synaptic}. 
The maximal capacity is essentially controlled by the ratio between the recovery time of the available
synaptic resources $\taud$ and the membrane time constant of the excitatory populations;
conversely it reveals a weaker dependence on the ratio between faciltation and depression time scales.
$N_c^{max}$ is also controlled by the excitatory and inhibitory drives.
As a matter of fact, for our parameters we obtained $N_c^{max} \simeq 4-5$, in pretty good
agreement with the measured maximal capacity.}

Furthermore, we observed that the power in the $\gamma$-band (25-100 Hz) increases with the number of loaded items $N_\mathrm{I}$,
in agreement with several experimental studies related to WM \cite{howard2003gamma,van2010hippocampal,roux2012gamma}.
\red{Interestingly, for the $\gamma$-rhythms, the inhibitory pool and the excitatory
populations contribute equally to its generation, confirming that its origing is related
to a PING-like mechanism. Instead the power in the $\beta$-band reveals a non monotonic behaviour with $N_\mathrm{I}$,
characterized by a rapid increase passing from 1 to 2 items, a small drop from 2 to 3 items
while remaining essentially constant for $N_\mathrm{I} \ge 3$. The activity in the $\theta$-band
is associated to the single excitatory population dynamics only, due to the fact that
the inhibitory population is not involved in the memory maintenance of single items.
No variation with $N_I$ have been observed in the $\alpha$-band, analogously
to what reported for experiments on humans during memory retention in \cite{tallon1998}.}
 
 \red{Finally, we have defined a measure of the memory capacity in terms of
 the mean membrane potential measured just after a PB deliverance. This quantity
 increases with the number of loaded items for $N_I < 5$ and saturates
 for $N_I \ge 5$. These results resemble the ones obtained for the
 ERPs detected in young adults performing visual memory tasks \cite{vogel2004neural,vogel2005neural}.
This analysis suggests that in our neural mass model the value of the mean membrane potential can be
employed to measure memory load and capacity, in analogy with the neurophysiological indicator
defined in \cite{vogel2004neural,vogel2005neural}.}

\red{However our neural mass model presents several simplifications from a biological point of view, such 
as the pulsatile interactions or the absence of transmission delays, therefore more realistic aspects should be included in future developments. 
As shown for this next generation of neural mass models,
in absence of plasticity, the inclusion of the time scales associated to the rise and decay
of the post-synaptic potentials could induce the emergence of new oscillatory rhythms \cite{devalle2017firing, coombes2019},
while the delayed synaptic transmission could lead to more complex macroscopic behavious
\cite{devalle2018}. For what concerns the plasticity terms, we expect to further improve the agreement 
between the neural mass model and the network dynamics with $\upmu$-STP by including
the correlations and the fluctuation of the microscopic synaptic variables in the mean-field
formulation, analogously to what done in \cite{schmutz2020}.}
 
A fundamental aspect of WM, not included in our model, is the volitional control, 
which is a cognitive function that allows for the control of behaviour from the environment and 
to turn it towards our internal goals \cite{goldman1995cellular}.
A flexible frequency control of cortical oscillations has been recently proposed as an unified mechanism for the rapid 
and controlled transitions of the WM between different computational operations 
\cite{dipoppa2013flexible,dipoppa2016,schmidt2018network}. In particular,
the authors consider as a neural correlate of WM a 
bistable network with coexisting persistent and resting states that an external
periodic modulation can drive from an operating mode to another one depending on the frequency 
of the forcing.  However, in \cite{schmidt2018network} it has been suggested that the forcing term,
in order to be considered more realistic, should self-emerge by the network dynamics in 
form of trains of periodic PBs and not be imposed from the exterior.
In our model for synaptic-based WM we have shown that some WM operations are associated with PBs delivered at different
frequencies: namely, item loading and recall with transient oscillations in the $\delta$ band, as in \cite{schmidt2018network},
joined to burst oscillations in $\beta-\gamma$ band as in \cite{dipoppa2013flexible};
multi-item maintenance to harmonics in the $\beta-\gamma$ band. Therefore, we believe that our
neural mass model with STP can represent a first building block for the development of an unified control mechanism
for WM, relying on the frequencies of deliverance of the self-emerging trains of PBs. However, a development towards 
realistic neural architectures would require to design a multi-layer network topology 
to reproduce the interactions among superficial and deep cortical layers \cite{miller2018working}.

%\clearpage
\section*{Methods}\label{sec:methods}

\subsection*{Spiking neuronal network model}
\label{sec:nn}

The membrane potential dynamics of the $i$-th  QIF neuron in a network of size $N$ can be written as 
\begin{align}\label{eq:qifnetwork1}
	\taum \dot{V}_i=V_i^2(t) +\eta_i +\Iback + \Istim(t)+ \taum \frac{1}{N} \sum_{j=1}^N {\tilde J}_{ij}(t) S_j(t) \quad,  \qquad i =1,\dots,N
\end{align} 
where $\taum = 15$ ms is the membrane time constant and ${\tilde J}_{ij}(t)$ the strength of the direct synapse
from neuron $j$ to $i$ that, in absence of plasticity, we assume to be constant and all identical, i.e.
${\tilde J}_{ij}(t) = J$. The sign of $J$ determines if the neuron is excitatory ($J >0$) or inhibitory ($J <0$).
Moreover,  $\eta_i$ represents the neuronal excitability,
$\Iback$ a constant background DC current, $\Istim (t)$ an external stimulus 
and the last term on the right hand side the synaptic current due to the
recurrent connections with the pre-synaptic neurons.
 For instantaneous post-synaptic potentials (corresponding to $\delta$-spikes) the 
 neural activity $S_j(t)$ of neuron $j$ reads as
\begin{align}
\label{eq:sfiringrate}
	 S_j(t) = \sum_{t_j(k) <t}\delta(t-t_j(k)),
\end{align}
where $S_j(t)$ is the spike train produced by the $j$-th neuron and $t_j(k)$ denotes the $k$-th spike time in such
sequence. We have considered a fully coupled network without autapses, therefore the post-synaptic current will be
the same for each neuron apart corrections of order ${\cal O}(1/N)$.

In the absence of synaptic input, external stimuli and $\Iback=0$, the QIF neuron exhibits two possible dynamics, depending on the sign of $\eta_i$.
For negative $\eta_i$, the neuron is excitable and for any initial condition $V_i(0)< \sqrt{-\eta_i}$, it reaches asymptotically the resting value $-\sqrt{-\eta_i}$. On the other hand, for initial 
values larger than the excitability threshold, $V_i(0)> \sqrt{-\eta_i}$, the membrane potential grows unbounded and a reset mechanism has to be introduced to describe the spiking behaviour of a neuron. Whenever $V_i(t)$ reaches a threshold value $\Vp$, the neuron $i$ delivers a spike and its membrane voltage is reset to $\Vr$, for the QIF neuron $\Vp=-\Vr = \infty$. For positive $\eta_i$ 
the neuron is supra-threshold and it delivers a regular train of spikes with frequency $\nu_0 = \sqrt{\eta_i}/\pi$.

As already mentioned, the analytic derivation of the neural mass model can be performed
when the excitabilities $\{\eta_i\}$ follows a Lorentzian distribution
\begin{align}
g(\eta)= \frac{1}{\pi} \frac{\Delta}{(\eta-H)^2 + \Delta^2}. 
\label{eq:lorentziang}
\end{align}
centred at $H$ and with HWHM $\Delta$. 
In particular, the excitability values $\eta_i$ have been generated deterministically for a network of $N$ neurons by
employing the following expression
\begin{align}\label{eq:detlorentzian}
	\eta_i=H+\Delta \tan \left[\frac{\pi}{2}\frac{2i-N-1}{N+1}\right] \quad .
\end{align}

Numerical simulations of the QIF networks have been performed  by employing an Euler scheme with a  timestep $\Delta t=10^{-4}\taum$.
Moreover, when performing the numerical experiments, analogous to the approach in \cite{montbrio2015macroscopic},
the threshold and reset values have been approximated to $V_p=-V_r = 100$ and a refractory period is introduced to deal with
this approximation. In more details, whenever the neuron $i$ reaches $V_i\geq 100$, its voltage is reset to $V_i=-100$ and the voltage remains unchanged at the reset value for a time interval $\frac{2\taum}{100}$, accounting for the time needed to reach $V_i = \infty$ from
$V_i = 100$ and to pass from  $V_i = -\infty$ to $V_i= -100$. The spike emission of neuron $i$ is registered at half 
of the refractory period. 

\subsection*{Short-term synaptic plasticity}
\label{sec:stp}

By following \cite{mongillo2008synaptic} in order to reproduce WM mechanisms in the PFC we assume that 
excitatory-to-excitatory synapses display depressing and facilitating transmission, described by a phenomenological model of STP developed by  Markram, Tsodyks and collaborators \cite{tsodyks1997neural,tsodyks1998neural,markram1998}. In this model, each pre-synaptic spike depletes the neurotransmitter supply and, at the same time, the concentration of intracellular calcium.
The accumulation of calcium ions is responsible for an increase of the release probability of neurotransmitters 
at the next spike emission. Each synapse displays both depression on a time-scale $\taud$, due to the neurotransmitter depletion,
and facilitation on a time-scale $\tauf$, linked to the increased calcium concentration and release probability.
Experimental  measurements for the facilitating excitatory synapses in the PFC indicates that $\tauf \simeq 1$ s and that $\tauf \gg \taud$ \cite{wang2006}.

As done in a previous analysis \cite{di2013}, we also assume for the sake of simplicity, 
that all the efferent synapses of a given neuron follow the same dynamical evolution, therefore
they can be characterized by the index of the pre-synaptic neuron. In particular,
the depression mechanism is mimicked by assuming that, at certain time $t$, each synapse has at disposal $X_i(t)$ synaptic resources 
and that each pre-synaptic spike uses a fraction $U_i(t)$ of these resources. In absence of emitted spikes
the synapses will recover the complete availability of their resources on a time scale $\taud$, i.e. $X_i(t \gg \taud) \to 1$.
The facilitation mechanism instead leads to an increase of the fraction $U_i(t)$ of employed resources at each spike by
a quantity $U_0(1-U_i)$. In between spike emissions $U_i$ will recover to its baseline value $U_0$ on a time scale $\tauf$.
The dynamics of the synaptic variables $X_i(t)$ and $U_i(t)$ is ruled by the following ordinary differential equations: 
\begin{subequations}\label{eq:du1}
	\begin{align}
	\dot{X}_i &=\frac{1-X_i(t)}{\taud}-X_i(t) U_i(t) S_i(t)\\
	\dot{U}_i &=\frac{U_0-U_i(t)}{\tauf}+U_0(1-U_i(t)) S_i(t)
	\enskip .
	\end{align}
\end{subequations} 
In this framework $U_i X_i$ represents the amount of resources employed to produce a post-synaptic potential,
therefore the synaptic coupling ${\tilde J}_{ij}(t)$ entering in Eq. \eqref{eq:qifnetwork1}
in presence of STP will be modified as follows:
\begin{equation}\label{eq:splastic}
{\tilde J}_{ij} (t) = J U_j(t) X_j(t)  \quad \forall i \quad .
\end{equation}
  
The spiking network model with microscopic STP ($\upmu$-STP) (Eqs. \eqref{eq:qifnetwork1}, \eqref{eq:du1} and \eqref{eq:splastic}) consists of $3N$ differential equations: one for the membrane voltage and two for the synaptic variables of each neuron. In order to reduce the system size effects and to obtain accurate simulations reproducing the neural mass dynamics, extremely large network sizes are required, namely $N > 100,000$. 
%\blue{In particular, this degree of accuracy is needed, since the parameter window around $\Iback=1.532$, in which single population %PBs and juggling can be observed (see \nameref{sec:competitiontwoitem}) is very narrow  and close to the critical value of $\Iback$  %at which a Hopf bifurcation is present (see \nameref{sec:bif}).}

For these network simulations massive numerical resources are required, however the complexity of the synaptic dynamics can be 
noticeably reduced, by treating the STP at a mesoscopic level under the assumption that the spike trains emitted by each neuron 
are Poissonian \cite{tsodyks1998neural,schmutz2020}. In this framework, the mesoscopic description of the synapse can be
written as 
\begin{subequations}\label{eq:seminetwork}
	\begin{align}
	\dot{x} &=\frac{1-x(t)}{\taud}-u(t) x(t) A(t)\label{eq:mfx}\\
	\dot{u} &=\frac{U_0-u(t)}{\tauf}+U_0(1-u(t)) A(t) \quad , \label{eq:mfu}
 	\end{align}
\end{subequations}
where $x= \langle X_i \rangle$ and $u=\langle U_i \rangle$ represent the average of the microscopic variables $\{X_i\}$ and $\{U_i \}$ over the whole population\footnote{The population average $\langle \cdot \rangle = \frac{1}{N_l} \sum_{i=1}^{N_l} \cdot $ is 
performed over all the neurons $N_l$ in population $l$} and $A(t) = \langle S_j(t) \rangle$ is the mean neural activity.
 
The synaptic coupling entering in Eq. \eqref{eq:qifnetwork1} now becomes:
\begin{align}\label{eq:splastic:mean}
{\tilde J}_{ij} (t) = J u(t) x(t)  \quad \forall i,j \quad .
\end{align}
The dynamics of the QIF network with the mesoscopic STP (m-STP) is given by  $N+2$ ODEs, namely Eqs. \eqref{eq:qifnetwork1} and \eqref{eq:seminetwork} with ${\tilde J}_{ij} (t)$ given by \eqref{eq:splastic:mean}, therefore we can reach larger system sizes than with the
$\upmu$-STP network model.

It should be noticed that the synaptic variables $X_i(t)$ and $U_i(t)$ for the same synapse 
are correlated, since they are both driven by the same spike train $S_i(t)$ delivered by neuron $i$.
These correlations are neglected in the derivation of the m-STP model, therefore one can write $\langle U_i(t) X_i(t) \rangle  =  \langle U_i(t) \rangle \langle X_i(t) \rangle = u(t) x(t) $. This approximation 
is justified in \cite{tsodyks1998neural} by the fact that
the coefficients of variation of the two variables  $U_i$ and $X_i$ are particularly small for facilitating synapses.
Indeed it is possible to write a stochastic mesoscopic model for the STP that includes the second order moments for  $U_i$ 
and $X_i$, i.e. their correlations and fluctuations \cite{schmutz2020}.
 
In our simulations we fixed $\taud=200$ ms, $\tauf=1500$ ms and $U=0.2$ and we integrated the network
equations by employing a standard Euler scheme with time step $0.0015$ ms.

\subsection*{Exact neural mass model with STP}
\label{sec:nm}
For the heterogeneous QIF network with instantaneous synapses (Eqs.~\eqref{eq:qifnetwork1}-\eqref{eq:sfiringrate}), 
an exact neural mass model, in absence of STP, has been derived  in \cite{montbrio2015macroscopic}. 
The analytic derivation is possible for QIF spiking networks thanks to the Ott-Antonsen Ansatz
\cite{ott2008} applicable to phase-oscillators networks whenever the natural frequencies (here corresponding to 
the excitabilities $\{ \eta_i \}$) are distributed accordingly to a Lorentzian distribution with median $H$ and HWHM $\Delta$.
In particular, this neural mass model allows for an exact macroscopic description of the population dynamics, in the thermodynamic limit $N\rightarrow \infty$, in terms of only two collective variables, namely the mean membrane voltage $v(t)$ and the instantaneous population rate $r(t)$, as follows
\begin{subequations}
\label{eq:physrevx}
	\begin{align}
		\taum\dot{r}(t)&=\frac{\Delta}{\taum \pi}+2r(t) v(t)\\
		\taum\dot{v}(t)&=v^2(t) + H +\Iback + \Istim(t) -\left[\pi \taum r(t)\right]^2 + \taum {\tilde J}(t) r(t) \quad ;
	\end{align}
\end{subequations}
where the synaptic strength is assumed to be identical for all neurons and for instantaneous synapses in absence of plasticity
${\tilde J}(t) = J$. However, by including a dynamical
evolution for the synapses and therefore additional collective variables, this neural mass model can be extended 
to any generic post-synaptic potentials, see e.g. \cite{devalle2017firing} for exponential synapses or
\cite{coombes2019} for conductance based synapses with $\alpha$-function profile.

In our case the synaptic evolution at a mesoscopic level is given by the m-STP model described above,
therefore it will be sufficient to include the dynamical evolution of the m-STP in the
model \eqref{eq:physrevx} to obtain an exact neural mass model for the QIF spiking network with STP.
In more details, we consider the equations \eqref{eq:seminetwork},\eqref{eq:splastic:mean} for the m-STP
dynamics and we substitute the population activity $A(t)$ with the instantaneous firing rate $r(t)$, that corresponds to
its coarse grained estimation in the limit $N \to \infty$. Furthermore, the synaptic coupling entering
in Eq. \eqref{eq:physrevx} will become ${\tilde J}(t) = J u(t) x(t)$ and the complete neural mass model
reads as 
\begin{subequations}\label{eq:mf1pop}
	\begin{align}
		\taum\dot{r}(t)&=\frac{\Delta}{\taum \pi}+2r(t)v(t)\\
		\taum\dot{v}(t)&=v^2(t) + H +\Iback + \Istim(t) -\left[\pi \taum r(t)\right]^2 + J\taum u(t)x(t)r(t)\\
		\dot{x}(t) &=\frac{1-x(t)}{\taud}-u(t)x(t)r(t)\\
		\dot{u}(t) &=\frac{U_0-u(t)}{\tauf}+U_0(1-u(t))r(t) \quad .
	\end{align}
\end{subequations}

The macroscopic dynamics generated by the neural mass model \eqref{eq:mf1pop} and by the QIF network
with $\upmu$-STP and m-STP are compared in the sub-section \nameref{sec:comparison}
of the section \nameref{sec:results}. In particular, in order to compare the value obtained from the simulation of the
neural mass model \eqref{eq:mf1pop} with those obtained by the integration of the microscopic networks, 
we evaluated the following instantaneous population average from the microscopic variables $v(t)=\langle V_i(t) \rangle$, 
$x(t)= \langle x_i \rangle$ and $u(t)=\langle  u_i \rangle$. 
To estimate the instantaneous population firing activity $r(t)$, appearing in \eqref{eq:mf1pop}, from 
the microscopic simulations is more complicated.
This estimation is based on the spike-count, namely one counts all the spikes $N_s(W)$ emitted in the network within a time window $W$. Then the firing rate is estimated as $r(t) = N_s(W)/W$ with $W= 0.01 \taum$. 
%Note that this estimation is used solely for the population firing rate presented in the figures. For the QIF network simulation, $v$, $x$ and $u$ are increased in an instantaneous manner whenever a spike is registered.
The neural mass model has been numerically integrated by employing the adaptive Dormand-Prince method
with a absolute and relative tolerances of $10^{-12}$.
 
\subsection*{Multi-populations models}
\label{sec:mp}

The discussed models can be easily extended to account for multiple interconnected neuronal populations. 
In the following we restrict this extension to the QIF network with m-STP and the corresponding neural mass model, since the simulations we performed for the QIF network with $\upmu$-STP are limited to a single population.

We consider a network composed of one inhibitory and $\Npop$ excitatory interacting neural populations, each composed by $N_k$ neurons. Therefore the dynamics of the membrane potential $V_{i,k}(t)$ of the $i$-th  QIF neuron of the $k$-th population ($k=0,\dots,\Npop$) and of the mesoscopic synaptic variables $(u_k(t),x_k(t))$ for the excitatory populations 
($k>0$) can be written as follows
\red{
\begin{subequations}\label{eq:mpopseminetwork}
	\begin{align}
	&\taum^{n} \dot{V}_{i,k} =V_{i,k}^2(t) +\eta_{i,k}+\Iback + \Istim^{(k)}(t) + \taum^{n}
	\sum_{l=0}^{\Npop} \frac{\tilde{J}_{kl}(t)}{N_l} \sum_{j(l)=1}^{N_l}  S_{j,l}(t) \\
	\nonumber
	 &\qquad \enskip i=1,\dots,N_k
	\\
	&\dot{x}_k =\frac{1-x_k(t)}{\taud}-u_k (t) x_k(t) A_k(t)\\
	&\dot{u}_k =\frac{U_0-u_k(t)}{\tauf}+U_0(1-u_k)  A_k(t)  \enskip ;
	\end{align}
\end{subequations}
}
where $\Istim^{(k)}(t)$ is the stimulation current applied to the population $k$ and 
$A_k(t)$ is the population activity of the $k$-th population. \red{The index 
$j$ identifies the neuron belongs to population $l$ composed by $N_l$ neurons.
We assume that the synaptic couplings $\tilde{J}_{kl}$ depend on the population indices $k$ and $l$ but
not on the neuron indices; moreover we assume that the neurons are globally coupled both at the intra- and inter-population level.
The synaptic couplings for excitatory-excitatory connections are plastic, therefore they can be written as 
\begin{equation}
{\tilde J}_{kl}(t) =  J_{kl} u_l (t) x_l(t) \enskip , 
\label{eq:Jmulti}
\end{equation}
while the expression for the synaptic coupling will be simply set to
$ {\tilde J}_{kl}(t) =  J_{kl} $ if one of the populations $k$ or $l$ is inhibitory.}
The sign is determined by the pre-synaptic population $l$, if it is excitatory (inhibitory)
$J_{kl}>0$ ($J_{kl} < 0$). In this paper we have always considered a single inhibitory
population  indexed as $k=0$  and $\Npop$ excitatory populations with positive indices,
and usually assumed $\taum^{n} =15$ ms for excitatory and inhibitory populations,
apart in sub-section \nameref{7items}  where we set  $\taum^{\mathrm{e}}=15$ ms and $\taum^{\mathrm{i}} =10$ ms. For each population, we always considered $N_k = 200,000$ neurons.

The corresponding multi-population neural mass model can be straightforwardly written as
\begin{subequations}\label{eq:mpopmeanfield}
	\begin{align}
		\taum^{n} \dot{r_k}&=\frac{\Delta_k}{\taum^{n} \pi} + 2 r_k(t) v_k(t)  \quad k=0,1,\dots,\Npop \\
		\taum^{n} \dot{v}_k&=v_k^2(t) + H_k+\Iback + \Istim^{(k)}(t)  -(\pi \taum^{n} r_k(t))^2 + \taum^{n} \sum_{l=0}^{\Npop} {\tilde J}_{kl}(t)r_l(t)\\
		\dot{x}_l &=\frac{1-x_l(t)}{\taud}-u_l(t) x_l(t) r_l(t)\\
		\dot{u}_l &=\frac{U_0-u_l(t)}{\tauf}+U_0(1-u_l(t))r_l(t)  \qquad l=1,\dots,\Npop \quad ;
	\end{align}
\end{subequations} 
where for excitatory-excitatory population interactions
\begin{equation}
{\tilde J}_{kl}(t) =  J_{kl} u_l (t) x_l(t) \quad ; 
\label{eq:Jmulti2}
\end{equation}
and whenever population $k$ or $l$ is inhibitory  $ {\tilde J}_{kl}(t) =  J_{kl} $.

\subsection*{Bifurcation analysis}
\label{sec:bif}

In order to exemplify the dynamical mechanisms underlying the maintenance of different items in presence 
of STP, we analyse the bifurcation diagram for the model \eqref{eq:mpopmeanfield} with three populations.
In particular, we study the emergence of the different dynamical states occurring in a network made of
two excitatory and one inhibitory population corresponding to the network architecture introduced in sub-section \nameref{sec:ark}.

\begin{figure}[!h]
%\begin{adjustwidth}{-2.25in}{0in}
	\includegraphics[width=1\linewidth]{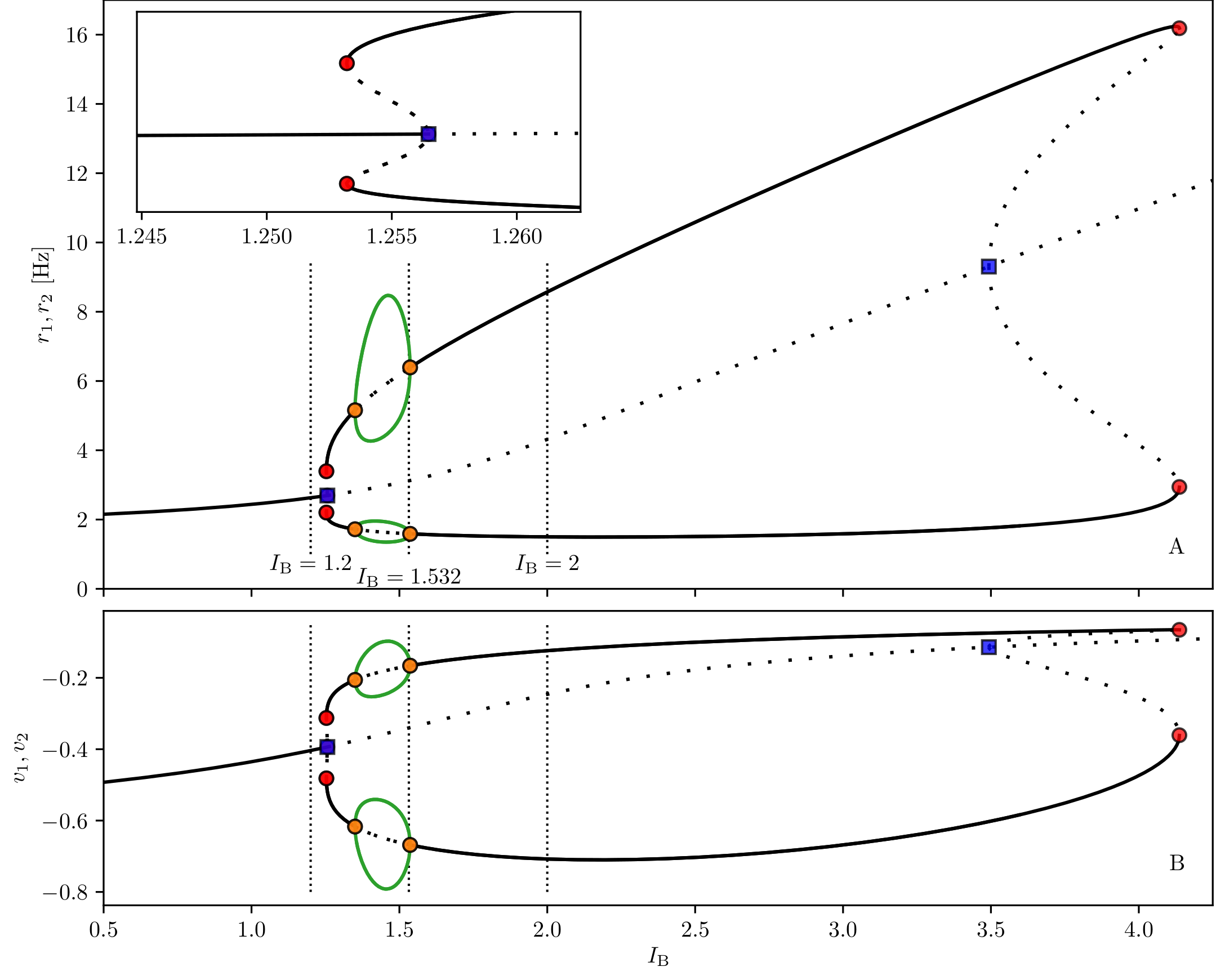}
	\caption{\textbf{Bifurcation diagram for two excitatory populations}.
	Bifurcation diagram displaying the instantaneous firing rate\red{s} $r_k$ \red{{\bf (A)} and mean membrane potentials $v_k$ {\bf (B)} }for the excitatory populations as a function of the background current $\Iback$. Solid (dashed) black lines refer to stable (unstable) asynchronous states, while green solid lines denote the
	maxima and the minima of stable collective oscillations. Symbol refer to bifurcation points:
	branch points (blue squares), Hopf bifurcations (orange circles) and saddle-node bifurcations (red circles).
	The inset in \red{(A)} displays an enlargement of the bifurcation diagram. \red{The lower (upper) branch of equilibria in (A) corresponds to the lower (upper) branch in (B)}.
	All the remaining parameters are as in Fig. \ref{fig0}. Continuation performed with the software AUTO-07P \cite{auto}.}
	\label{bifurcation_diag}
%\end{adjustwidth}
\end{figure}

Due to the symmetries in the synaptic couplings and in the structure of the dynamical equations  \eqref{eq:mpopmeanfield}
the macroscopic dynamics observable for the two excitatory networks will be equivalent. Therefore,
we will display the bifurcation diagram in terms \red{of the instantaneous firing rate $r_k(t)$ and the mean membrane potential $v_k(t)$ of one of the two excitatory populations ($k >0$)} as a function of the common background current $\Iback$ by fixing all the other parameters of the model \eqref{eq:mpopmeanfield} as in Fig. \ref{fig0}.
The phase diagram, shown in Fig. \ref{bifurcation_diag}, reveals that at low values of the background current ($\Iback\leq 
I_{\rm{sn}}^{(1)} \simeq 1.2532$) there is a single stable fixed point with an asynchronous low firing dynamics.
This looses stability at $I_{\rm{bp}}^{(1)} \simeq 1.25647 $ giving rise to two coexisting stable fixed points with asynchronous dynamics: one at low firing rate due to spontaneous activity in the network and one at high firing rate corresponding to a persistent state. As shown in the inset in Fig. \ref{bifurcation_diag} (A), there is a small region $\Iback \in [I_{\rm{sn}}^{(1)},I_{\rm{bp}}^{(1)}]$ where we can have the coexistence of these three stable asynchronous states.

At $\Iback \equiv I_{\rm hb}^{(1)} \simeq 1.34998 $ we observe the emergence of coexisting collective oscillations
(periodic PBs) at low and high firing rates via supercritical Hopf bifurcations.
These oscillations exist in a quite limited range of parameters, namely $\Iback \in [I_{\rm hb}^{(1)},I_{\rm hb}^{(2)}]$, and they disappear at  $I_{{\rm hb}}^{(2)} \simeq 1.5363$.
Beyond this parameter value we again have a persistent state coexisting with a low firing activity regime,
these states finally annihilate with two unstable fixed point branches at $ I_{{\rm sn}}^{(2)} \equiv 4.13715$.

The knowledge of the bifurcation diagram shown in Fig. \ref{bifurcation_diag} allows us
to interpret the numerical experiments discussed in sub-section \nameref{sec:maint} and
displayed in Fig. \ref{fig0}.

Let us consider the first experiment, reported in column (A) of Fig. \ref{fig0} and showing 
the selective reactivation of the WM item via a nonspecific read-out signal. The item is firstly loaded 
into population one via a specific step current of amplitude $\Delta I^{(1)}(t)=0.2$ 
for a time interval $\Delta T_1 = 350$ ms. The selective reactivation of the target is obtained by applying 
a non-specific readout signal of amplitude $\Delta I^{(1)}(t)=\Delta I^{(2)}(t)=0.1$ to all excitatory neurons in both populations for a shorter time interval, namely $\Delta T_1 = \Delta T_2 = 250$ ms. For this numerical experiment we fixed 
$\Iback= 1.2 < I_{{\rm sn}}^{(1)} $, thus the only possible stable regime is a low firing asynchronous activity.
The item is loaded into population one by increasing, for a limited time window, the background current only for this population  
to the level $\Iback + \Delta I^{(1)} > I_{\rm hb}^{(1)}$, thus leading to the emission of a series of PBs,
whose final effect is to strongly facilitate the efferent synapses of population one. 
The sub-sequent application of a non-specific read-out signal amounts to an effective increase 
of their common background current to a value beyond $I_{\rm bp}^{(1)}$, where the persistent state coexists with the low
firing activity. Indeed during the read-out stimulation, population one displays a burst of high activity,
due to the facilitated state of its synapses, while population two is essentially unaffected by the read-out signal.
As soon as the stimulus is removed, the system moves back to the spontaneous activity regime.

The second experiment, shown in the column (B) of Fig. \ref{fig0}, concerns spontaneous reactivation of the WM item
via collective oscillations (periodic PBs). In this case the background current is set to $\Iback=1.532$, 
in order to have the coexistence of two stable limit cycles corresponding to periodic PBs with low and high firing rate.
The whole system is initialized in the asynchronous regime with spontaneous activity (which is unstable for this value
of $\Iback$); upon the presentation of the stimulus to population one, this jumps  to the upper limit cycle after loading the item into the memory. Once the stimulation is removed, due to the periodic synaptic refreshment, population one reactivates spontaneously 
by emitting a periodic sequence of PBs that is terminated by reducing $\Iback$ to a value smaller than $I_{{\rm sn}}^{(1)}$.

The last experiment shown in column (C) of Fig. \ref{fig0} refers to the spontaneous reactivation of the memory 
associated to a persistent state activity. In this case we set $\Iback=2$, thus the system is in an asynchronous regime 
beyond $I_{{\rm hb}}^{(2)}$, where there is coexistence of persistent and low firing activity beyond $I_{{\rm hb}}^{(2)}$. 
As in the previous experiment, the system is initialized in the asynchronous unstable regime. Upon a brief stimulation, population
one is led in the high activity persistent regime. Reducing the background current to $\Iback=1.2$ stops the persistent
activity. 

\red{
\subsection*{Heuristic Firing Rate Model}
\label{sec:wilsoncowan}
Firing rate models have been developed to describe heuristically the
dynamics of a neuronal population in terms of the associated firing rate $r$;
one of the most known example is represented by the Wilson-Cowan model \cite{wilson1972excitatory}.
These models are usually written as  \cite{dayan2001}
\begin{equation}\label{eq:wc1}
	\taum \dot{r}=-r+\Phi(I)
	\enskip ,
\end{equation}
where $I$ represents the total input current received by each neuron in the population and
$\Phi(I)$ is the steady-state  firing rate solution, or activation function.
This function is usually assumed to be a sigmoid function and it is determined on
the basis of the dynamical features of the neurons in the considered population.
As stated in \cite{devalle2017firing}, these firing rate models, despite being extremely
useful to model brain dynamics, do not take into account synchronization phenomena
induced by the sub-threshold voltage dynamics. Therefore, these
firing rate models fail in reproducing fast oscillations observed in inhibitory networks, 
without the addition in their dynamics of an ad-hoc time delay. 
These collective oscillations are instead captured by the neural mass model
introduced in \cite{montbrio2015macroscopic} and considered in this paper.}

\red{By following the analysis in \cite{devalle2017firing}, we can obtain a firing rate model 
corresponding to the exact neural mass ODEs \eqref{eq:physrevx}. More specifically this heuristic firing rate can be derived for a QIF network of spiking neurons,
by considering the corresponding steady-state solution $(v^\ast,r^\ast)$ given by
\begin{subequations}
\label{eq:sfi1}
	\begin{align}
		0&=\frac{\Delta}{\taum \pi}+2r^\ast v^\ast\\
		0&=(v^\ast)^2- (\pi \taum r^\ast)^2  + I
	\end{align}
\end{subequations}
where $I=H +\Iback + \Istim + \taum {\tilde J} r^\ast$. This leads to
a self-consistent equation for the steady-state firing rate, given
by  $r^\ast=\Phi(I)$, where 
\begin{align}\label{eq:sfi2}
	\Phi(I)=\frac{1}{\sqrt{2}\pi \taum}\sqrt{I+\sqrt{I^2+\Delta}}\quad .
\end{align}
The equations \eqref{eq:wc1} and \eqref{eq:sfi2} represent a firing rate model 
corresponding to the QIF spiking network with m-STP. 
This firing rate model has been considered in sub-section \nameref{comparison_fr} 
in order to perform numerical experiments on WM maintenance (see Fig. \ref{fig:wilsoncowan}). 
}

\subsection*{Maximal Working Memory Capacity}
\label{sec:capacity}

By following \cite{mi2017synaptic} we can give an estimate of the maximal
memory capacity for our neural mass model \eqref{eq:fre}  with m-STP  \eqref{eq:dumeanfied}.
The maximal capacity can be estimated as the ratio of two time intervals
\begin{equation}
N_\mathrm{c}^\mathrm{max} \simeq \frac{T_\mathrm{c}^\mathrm{max}}{T_\mathrm{b}}
\label{maxmem}
\end{equation}
where $T_\mathrm{c}^\mathrm{max}$ is the maximal period of the network limit cycle and $T_\mathrm{b}$
the inter-burst interval between two successive PBs.
In \cite{mi2017synaptic}, $T_\mathrm{c}^\mathrm{max}$ has been estimated as the time needed to the synaptic efficacy $u_k(t) x_k(t)$
of a generic population $k$ to recover to the maximum value after a PB emission. Since all
the excitatory populations are identically connected among them and with the inhibitory population,
this time does not depend on the considered population. The approximate expression reported in \cite{mi2017synaptic} is the following
\begin{equation}
T_\mathrm{c}^\mathrm{max} \simeq \taud \ln{\frac{\tauf/\taud}{1-U_0}}
\label{tcmax}
\end{equation}
As expected the recovery time is essentially ruled by the depression time scale $\taud$.

It can be shown that $T_\mathrm{b}$ has three components, i.e. the time width of the previous excitatory PB, 
the delay of the inhibitory burst triggered by the excitatory PB, plus its time width, and the  
time needed for the next active excitatory population  to recover from inhibition and to elicit
a PB. In our model framework, we can neglect the first two time intervals
and limit ourselves to estimate the latter time.

Let us denote the next firing population as the $m$-th one; we can assume that during $T_\mathrm{b}$ the 
connection strength does not vary much and that the firing rates are essentially
constants. Therefore we can rewrite the time evolution of the mean membrane potential
appearing in \eqref{eq:fre} as follows : 
\begin{equation}
\label{eq:fre2}
\taum^{\rm e} \dot{v}_k = v_k^2 + \left[ H^{(\mathrm{e})} -(\pi \taum^{\rm e} {\bar r}^{(\mathrm{e})})^2 
+\Iback + \taum^{\rm e} \left( - |J_{\rm ei}| {\bar r}^{(\mathrm{i})} + {\bar J} {\bar r}^{(\mathrm{e})} \right)  \right] = v_k^2 + C
\end{equation} 
where ${\bar J} = [\Jeep + (\Npop-2)\Jeeb] {\bar x} {\bar u} $ takes in account the synaptic 
efficacy of all excitatory synapses in an effective manner,
${\bar r}^{(\mathrm{i})}$ and ${\bar r}^{(\mathrm{e})}$ are the inhibitory and excitatory population rates
and $C$ is the constant quantity within square brackets on the right-hand side. 
The expression of $C$ can be further simplified by noticing that the quadratic term  is negligible
and by assuming that the excitatory and inhibitory firing rates are similar. Moreover, by assuming that
the excitatory neurons are almost uncoupled during $T_\mathrm{b}$, one gets:
\begin{equation}
\label{eq:C}
C = \left[H^{(\mathrm{e})} +\Iback + \taum^{\rm e} \left(-|J_{\rm ei}|+ {\bar J} \right) \frac{\sqrt{H^{(\mathrm{e})} + \Iback}}{\pi} \right] 
\end{equation} 
where ${\bar r}^{(\mathrm{i})}= {\bar r}^{(\mathrm{e})}$ and ${\bar r}^{(\mathrm{e})}= \frac{\sqrt{H^{(\mathrm{e})} + \Iback}}{\pi}$ as for an isolated QIF neuron driven
by the mean excitability and by the background current.

Therefore the time needed to the mean membrane potential to go from an initial negative value
$v_m(0) = V_0$, determined by the discharge of the inhibitory neurons, to a {\it threshold} value $V_{th}$, where the 
PB starts to be delivered, is given by
\begin{equation}
\label{eq:tb}
T_\mathrm{b}= \frac{\taum^{\rm e}}{\sqrt{C}} \left[ \arctan{\frac{V_{\mathrm{th}}}{\sqrt{C}}} - \arctan{\frac{V_0}{\sqrt{C}}} \right] \simeq \frac{\taum^{\rm e} \pi}{\sqrt{C}}
\end{equation} 
where on the right hand side of the equation we have finally assumed that $V_{th} >> 1 $ and $V_0 << -1$.

Thus the following expression for the maximal capacity is obtained
\begin{equation}
N_\mathrm{c}^{\rm max} \simeq  \frac{\taud}{\taum^{\rm e}} \ln{\left[\frac{\tauf/\taud}{1-U_0}\right]} \frac{\sqrt{C}}{\pi}
\quad .
\label{maxmem0}
\end{equation}

For the parameters employed in sub-section \nameref{7items} we obtained the following
theoretical values $T_\mathrm{c}^{\rm max} \simeq 447$ ms, $T_\mathrm{b} \simeq 93-126$ ms depending on the value of
$ 0.5 \le {\bar x} {\bar u} \le 1.0$, thus $3.6 \le N_\mathrm{c}^{\rm max} \le 4.8$ not far from the
measured value that was $N_\mathrm{c}^{\rm max} = 5$.

\subsection*{Spectrogram Estimation}
\label{sec:spect}

In order to generate the spectrograms shown in Figs. \ref{fig0}, \ref{fig:wilsoncowan}, \ref{fig:two_item_read_out}, \ref{fig:7item_setup_load3}, \ref{fig:7item_setup_load6} and \nameref{Sfig2}, the \textit{signal} package, which is part of the SciPy library \cite{scipy}, is used. The subroutine \textit{stft} (short time Fourier transform, STFT) generates Fourier transforms $\mathcal{F}[s(t)](t,f)$ of a signal $s(t)$ within a running time window of length $\Delta T_\mathrm{win}$ at time $t$. The STFT is performed using overlapping windows (95\% overlap) throughout this work. For Figs. \ref{fig0}, \ref{fig:wilsoncowan} and \ref{fig:two_item_read_out} the window length is set to $\Delta T_\mathrm{win}=0.2$ s, leading to a sufficiently fine resolution in time and frequency.  For Figs. \ref{fig:7item_setup_load3},  \ref{fig:7item_setup_load6} and \nameref{Sfig2} it is set to $\Delta T_\mathrm{win}=1$ s, resulting in a better frequency resolution and decrease of time-resolution. The colours in the spectrograms code the normalized power spectral density $|\mathcal{F}[v_k(t)](t,f)|^2/(\max |\mathcal{F}[v_k(t)](t,f)|^2)$ obtained from voltage signals $v_k$ of different populations. For better visibility a log10 scale is used and values $<10^{-2}$ set to $10^{-2}$. 

\red{Since the average membrane potential is not accessible for the firing rate model \eqref{eq:wc1},\eqref{eq:sfi2}, in this case we made use of simulated local field potentials $\mathrm{LFP}_k$ in order to estimate the spectrograms. By following \cite{mazzoniEncoding2008}, we have estimated the 
local field potentials for the three populations appearing in the multi-item architecture displayed in Fig. \ref{fig:topology}
 as the sum of the absolute values of the synaptic inputs stimulating each populations:
\begin{subequations}
\label{LFP}
	\begin{align}
		\mathrm{LFP}_0 &= -[|J_{\rm ie}|(r_1+r_2)+ |J_{\rm ii}|r_0]\\
		\mathrm{LFP}_1 &= -[|\Jeep|x_1 u_1 r_1+|\Jeeb|x_2 u_ 2r_2+ |J_{\rm ei}|r_0]\\
		\mathrm{LFP}_2 &= -[|\Jeep|x_2 u_2 r_2 +|\Jeeb|x_1 u_1 r_1+ |J_{\rm ei}|r_0]
	\end{align}
\end{subequations}
where we neglect the constant current components for the calculation of the LFPs, as they do not contribute to the frequency spectra. Furthermore, to make possible a comparison with experimental measurements where high (low) activity states correspond to a minimum (maximum) value of the LFPs, 
we reversed the sign of the synaptic inputs in \eqref{LFP}. 
}

%\section*{Funding}
%AT received financial support by the Excellence Initiative I-Site Paris Seine (Grant No ANR-16-IDEX-008), by the Labex MME-DII (Grant No ANR-11-LBX-0023-01) and by the ANR Project ERMUNDY (Grant No ANR-18-CE37-0014), all part of the French programme ``Investissements d'Avenir''.

%\section*{Acknowledgements}
%The authors acknowledge preliminary discussions with D. Avitabile, M. Desroches, and E. Montbri\'o.

%\section*{Author Contributions}
%HT performed the simulations and data analysis.
%SO and AT were responsible for the state-of-the-art review and the paper write-up.
%All the authors conceived and planned the research.

%\section*{Conflict of Interest Statement}
%The authors declare that the research was conducted in the absence of any commercial or financial relationships that could be construed as a potential conflict of interest.

%\showthe\font

\clearpage
\section*{Supporting information}

%\paragraph*{Fig. S1}
%\begin{adjustwidth}{-2.25in}{0in}
%\includegraphics[width=1\linewidth]{../figures/persistent_state_switching.pdf}
%\label{Sfig1}
%{\bf Memory item switching with persistent activity}. The three columns (A-C) refer to three values of $\Delta T_2$ for which we
%have a memory switching from one item to the other for a background current $\Iback=2$ supporting persistent state activity.
%(A) $\Delta T_2=70$ ms; (B) $\Delta T_2=130$ ms and (C) $\Delta T_2=850$ ms. 
%Profiles of the stimulation current $\Istim^{(2)}(t)$ for the second excitatory population  ({\bf A1-C1}). Population firing rates $r_l$(t) ({\bf A2-C2}), normalized available resources 
%${\tilde x}_l(t)$ ({\bf A3-C3}) and normalised utilization factors ${\tilde u}_l(t)$ ({\bf A4-C4}) of the excitatory populations calculated from the simulations of the neural mass model. Other parameters as in Fig. \ref{fig0}.
%\end{adjustwidth}

\paragraph*{Fig. S1}

\begin{adjustwidth}{-1.25in}{0in}
\includegraphics[width=1\linewidth]{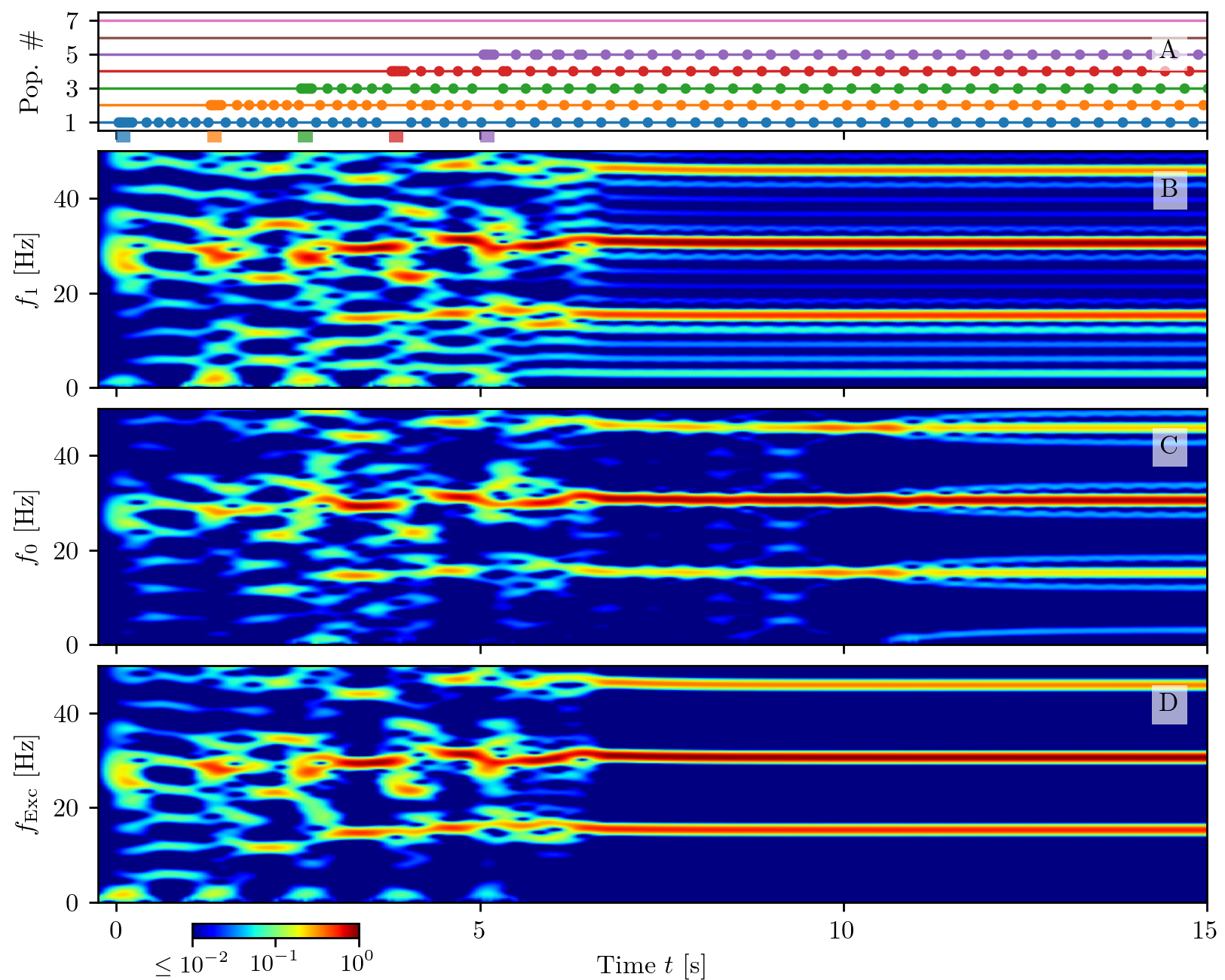}
\red{{\bf Maximal capacity.} Response of the system when $N_L=5$ excitatory populations are subsequentely stimulated at a presentation rate of 0.8 Hz. 	Population bursts of excitatory populations ({\bf A}): horizontal lines in absence of dots indicate quiescence phases at low firing rates $r_k$ for populations $k=1,...7$. Dots mark PBs of the corresponding population. The coloured bars on the time axis mark the starting and ending time of stimulating pulses, targeting each population. 
Spectrograms of the mean membrane potential $v_1(t)$ ({\bf B}), $v_0(t)$ ({\bf C}) and of the mean membrane potentials averaged over all the excitatory populations ({\bf D}); the corresponding frequencies have been denoted as $f_1$, $f_0$ and $f_\mathrm{Exc}$. Parameter values as in Fig. \ref{fig:7item_setup_load3}.}
\label{Sfig2}
\end{adjustwidth}

%\clearpage
\clearpage

%\paragraph*{Fig. S3}
%\begin{adjustwidth}{-1.25in}{0in}
%\includegraphics[width=1\linewidth]{../figures/7item_setup_load7_spectrogram.png} 
%\label{Sfig3}
%{\bf Multi-item memory loading: 7 items} Population spikes of excitatory populations: horizontal lines in absence of dots indicate quiescence phases at low firing rates $r_k$ for populations $k=1,...7$. Dots mark population spikes of the corresponding population. The coloured bars on the time axis mark the starting and ending time of stimulating pulses, targeting each population. 
%Spectrograms of the mean membrane potential $v_1(t)$ ({\bf B}), $v_0(t)$ ({\bf C}) and of the mean membrane potential
%averaged over all the excitatory populations ({\bf D}). Parameters as in Fig. \ref{fig:7item_setup_load3}.
%\end{adjustwidth}
\nolinenumbers
%\begin{thebibliography}{10}
 
%\bibitem{bib1}
%Conant GC, Wolfe KH.
%\newblock {{T}urning a hobby into a job: how duplicated genes find new functions}.
%\newblock Nat Rev Genet. 2008 Dec;9(12):938--950.

%\bibitem{bib2}
%Ohno S.
%\newblock Evolution by gene duplication.
%\newblock London: George Alien \& Unwin Ltd. Berlin, Heidelberg and New York: Springer-Verlag.; 1970.

%\bibitem{bib3}
%Magwire MM, Bayer F, Webster CL, Cao C, Jiggins FM.
%\newblock {{S}uccessive increases in the resistance of {D}rosophila to viral infection through a transposon insertion followed by a {D}uplication}.
%\newblock PLoS Genet. 2011 Oct;7(10):e1002337.

\bibliographystyle{plos2015} 
\bibliography{ref}

%\end{thebibliography}

\end{document}